\documentclass[twocolumn]{aastex631}

\shorttitle{Fine-structure constant with DESI}
\shortauthors{Jiang et al.}

\begin{document}

\title{Constraints on the spacetime variation of the fine-structure constant using DESI emission-line galaxies}

\correspondingauthor{Linhua Jiang}
\email{linhua.jiang@pku.edu.cn}

\author[0000-0003-4176-6486]{Linhua Jiang}
\affiliation{Department of Astronomy, School of Physics, Peking University, Beijing 100871, China}
\affiliation{Kavli Institute for Astronomy and Astrophysics, Peking University, Beijing 100871, China}

\author[0000-0003-0230-6436]{Zhiwei Pan}
\affiliation{Department of Astronomy, School of Physics, Peking University, Beijing 100871, China}
\affiliation{Kavli Institute for Astronomy and Astrophysics, Peking University, Beijing 100871, China}

\author{Jessica Nicole Aguilar}
\affiliation{Lawrence Berkeley National Laboratory, 1 Cyclotron Road, Berkeley, CA 94720, USA}

\author[0000-0001-6098-7247]{Steven Ahlen}
\affiliation{Physics Dept., Boston University, 590 Commonwealth Avenue, Boston, MA 02215, USA}

\author[0000-0002-8622-4237]{Robert Blum}
\affiliation{NSF NOIRLab, 950 N. Cherry Ave., Tucson, AZ 85719, USA}

\author{David Brooks}
\affiliation{Department of Physics \& Astronomy, University College London, Gower Street, London, WC1E 6BT, UK}

\author{Todd Claybaugh}
\affiliation{Lawrence Berkeley National Laboratory, 1 Cyclotron Road, Berkeley, CA 94720, USA}

\author[0000-0002-1769-1640]{Axel  de la Macorra}
\affiliation{Instituto de F\'{\i}sica, Universidad Nacional Aut\'{o}noma de M\'{e}xico,  Cd. de M\'{e}xico  C.P. 04510,  M\'{e}xico}

\author[0000-0002-4928-4003]{Arjun Dey}
\affiliation{NSF NOIRLab, 950 N. Cherry Ave., Tucson, AZ 85719, USA}

\author{Peter Doel}
\affiliation{Department of Physics \& Astronomy, University College London, Gower Street, London, WC1E 6BT, UK}

\author[0000-0003-2371-3356]{Kevin Fanning}
\affiliation{Kavli Institute for Particle Astrophysics and Cosmology, Stanford University, Menlo Park, CA 94305, USA}
\affiliation{SLAC National Accelerator Laboratory, Menlo Park, CA 94305, USA}

\author[0000-0003-4992-7854]{Simone Ferraro}
\affiliation{Lawrence Berkeley National Laboratory, 1 Cyclotron Road, Berkeley, CA 94720, USA}
\affiliation{University of California, Berkeley, 110 Sproul Hall \#5800 Berkeley, CA 94720, USA}

\author[0000-0002-2890-3725]{Jaime E. Forero-Romero}
\affiliation{Departamento de F\'isica, Universidad de los Andes, Cra. 1 No. 18A-10, Edificio Ip, CP 111711, Bogot\'a, Colombia}
\affiliation{Observatorio Astron\'omico, Universidad de los Andes, Cra. 1 No. 18A-10, Edificio H, CP 111711 Bogot\'a, Colombia}

\author{Enrique Gaztañaga}
\affiliation{Institut d'Estudis Espacials de Catalunya (IEEC), 08034 Barcelona, Spain}
\affiliation{Institute of Cosmology and Gravitation, University of Portsmouth, Dennis Sciama Building, Portsmouth, PO1 3FX, UK}
\affiliation{Institute of Space Sciences, ICE-CSIC, Campus UAB, Carrer de Can Magrans s/n, 08913 Bellaterra, Barcelona, Spain}

\author[0000-0003-3142-233X]{Satya  Gontcho A Gontcho}
\affiliation{Lawrence Berkeley National Laboratory, 1 Cyclotron Road, Berkeley, CA 94720, USA}

\author{Gaston Gutierrez}
\affiliation{Fermi National Accelerator Laboratory, PO Box 500, Batavia, IL 60510, USA}

\author{Klaus Honscheid}
\affiliation{Center for Cosmology and AstroParticle Physics, The Ohio State University, 191 West Woodruff Avenue, Columbus, OH 43210, USA}
\affiliation{Department of Physics, The Ohio State University, 191 West Woodruff Avenue, Columbus, OH 43210, USA}
\affiliation{The Ohio State University, Columbus, 43210 OH, USA}

\author{Stephanie Juneau}
\affiliation{NSF NOIRLab, 950 N. Cherry Ave., Tucson, AZ 85719, USA}

\author[0000-0003-1838-8528]{Martin Landriau}
\affiliation{Lawrence Berkeley National Laboratory, 1 Cyclotron Road, Berkeley, CA 94720, USA}

\author[0000-0001-7178-8868]{Laurent Le Guillou}
\affiliation{Sorbonne Universit\'{e}, CNRS/IN2P3, Laboratoire de Physique Nucl\'{e}aire et de Hautes Energies (LPNHE), FR-75005 Paris, France}

\author[0000-0003-1887-1018]{Michael Levi}
\affiliation{Lawrence Berkeley National Laboratory, 1 Cyclotron Road, Berkeley, CA 94720, USA}

\author[0000-0003-4962-8934]{Marc Manera}
\affiliation{Departament de F\'{i}sica, Serra H\'{u}nter, Universitat Aut\`{o}noma de Barcelona, 08193 Bellaterra (Barcelona), Spain}
\affiliation{Institut de F\'{i}sica d’Altes Energies (IFAE), The Barcelona Institute of Science and Technology, Campus UAB, 08193 Bellaterra Barcelona, Spain}

\author{Ramon Miquel}
\affiliation{Instituci\'{o} Catalana de Recerca i Estudis Avan\c{c}ats, Passeig de Llu\'{\i}s Companys, 23, 08010 Barcelona, Spain}
\affiliation{Institut de F\'{i}sica d’Altes Energies (IFAE), The Barcelona Institute of Science and Technology, Campus UAB, 08193 Bellaterra Barcelona, Spain}

\author[0000-0002-2733-4559]{John Moustakas}
\affiliation{Department of Physics and Astronomy, Siena College, 515 Loudon Road, Loudonville, NY 12211, USA}

\author{Eva-Maria Mueller}
\affiliation{Department of Physics and Astronomy, University of Sussex, Brighton BN1 9QH, U.K}

\author{Andrea  Muñoz-Gutiérrez}
\affiliation{Instituto de F\'{\i}sica, Universidad Nacional Aut\'{o}noma de M\'{e}xico,  Cd. de M\'{e}xico  C.P. 04510,  M\'{e}xico}

\author{Adam Myers}
\affiliation{Department of Physics \& Astronomy, University  of Wyoming, 1000 E. University, Dept.~3905, Laramie, WY 82071, USA}

\author[0000-0001-6590-8122]{Jundan Nie}
\affiliation{National Astronomical Observatories, Chinese Academy of Sciences, A20 Datun Rd., Chaoyang District, Beijing, 100012, P.R. China}

\author[0000-0002-1544-8946]{Gustavo Niz}
\affiliation{Departamento de F\'{i}sica, Universidad de Guanajuato - DCI, C.P. 37150, Leon, Guanajuato, M\'{e}xico}
\affiliation{Instituto Avanzado de Cosmolog\'{\i}a A.~C., San Marcos 11 - Atenas 202. Magdalena Contreras, 10720. Ciudad de M\'{e}xico, M\'{e}xico}

\author{Claire Poppett}
\affiliation{Lawrence Berkeley National Laboratory, 1 Cyclotron Road, Berkeley, CA 94720, USA}
\affiliation{Space Sciences Laboratory, University of California, Berkeley, 7 Gauss Way, Berkeley, CA  94720, USA}
\affiliation{University of California, Berkeley, 110 Sproul Hall \#5800 Berkeley, CA 94720, USA}

\author[0000-0001-7145-8674]{Francisco Prada}
\affiliation{Instituto de Astrof\'{i}sica de Andaluc\'{i}a (CSIC), Glorieta de la Astronom\'{i}a, s/n, E-18008 Granada, Spain}

\author[0000-0001-5589-7116]{Mehdi Rezaie}
\affiliation{Department of Physics, Kansas State University, 116 Cardwell Hall, Manhattan, KS 66506, USA}

\author{Graziano Rossi}
\affiliation{Department of Physics and Astronomy, Sejong University, Seoul, 143-747, Korea}

\author[0000-0002-9646-8198]{Eusebio Sanchez}
\affiliation{CIEMAT, Avenida Complutense 40, E-28040 Madrid, Spain}

\author[0000-0002-3569-7421]{Edward Schlafly}
\affiliation{Space Telescope Science Institute, 3700 San Martin Drive, Baltimore, MD 21218, USA}

\author{Michael Schubnell}
\affiliation{Department of Physics, University of Michigan, Ann Arbor, MI 48109, USA}
\affiliation{University of Michigan, Ann Arbor, MI 48109, USA}

\author[0000-0002-6588-3508]{Hee-Jong Seo}
\affiliation{Department of Physics \& Astronomy, Ohio University, Athens, OH 45701, USA}

\author{David Sprayberry}
\affiliation{NSF NOIRLab, 950 N. Cherry Ave., Tucson, AZ 85719, USA}

\author[0000-0003-1704-0781]{Gregory Tarlé}
\affiliation{University of Michigan, Ann Arbor, MI 48109, USA}

\author{Benjamin Alan Weaver}
\affiliation{NSF NOIRLab, 950 N. Cherry Ave., Tucson, AZ 85719, USA}

\author[0000-0002-6684-3997]{Hu Zou}
\affiliation{National Astronomical Observatories, Chinese Academy of Sciences, A20 Datun Rd., Chaoyang District, Beijing, 100012, P.R. China}

\collaboration{1}{The DESI Collaboration}

\begin{abstract}

We present strong constraints on the spacetime variation of the fine-structure constant $\alpha$ using the Dark Energy Spectroscopic Instrument (DESI). In this pilot work, we utilize $\sim110,000$ galaxies with strong and narrow [\ion{O}{3}] $\lambda\lambda$4959,5007 emission lines to measure the relative variation $\Delta\alpha/\alpha$ in space and time. The [\ion{O}{3}] doublet is arguably the best choice for this purpose owing to its wide wavelength separation between the two lines and its strong emission in many galaxies. Our galaxy sample spans a redshift range of $0<z<0.95$, covering half of all cosmic time. We divide the sample into subsamples in 10 redshift bins ($\Delta z=0.1$), and calculate $\Delta\alpha/\alpha$ for the individual subsamples. The uncertainties of the measured $\Delta\alpha/\alpha$ are roughly between $2\times10^{-6}$ and $2\times10^{-5}$. We find an apparent $\alpha$ variation with redshift at a level of $\Delta\alpha/\alpha=(2\sim3)\times10^{-5}$. This is highly likely to be caused by systematics associated with wavelength calibration, since such small systematics can be caused by a wavelength distortion of $0.002-0.003$ \AA, which is beyond the accuracy that the current DESI data can achieve. We refine the wavelength calibration using sky lines for a small fraction of the galaxies, but it does not change our main results. We further probe the spatial variation of $\alpha$ in small redshift ranges, and do not find obvious, large-scale structures in the spatial distribution of $\Delta\alpha/\alpha$. As DESI is ongoing, we will include more galaxies, and by improving the wavelength calibration, we expect to obtain a better constraint that is comparable to the strongest current constraint. 

\end{abstract}


\section{Introduction} \label{sec:intro}

The Standard Model of particle physics assumes that fundamental physical constants are universal and constant. They do not vary in space and time and their values can only be obtained from experiments. On the other hand, modern theories beyond the Standard Model predict or even require the variation of the fundamental constants \citep{Martins2017}. In the past several decades different methods, including laboratory experiments and astrophysical observations, have been developed to search for possible variations of these constants \citep[see][for a review]{Uzan2003,Uzan2011}. A particularly interesting constant is the fine-structure constant, denoted by $\alpha \equiv e^2 / \hbar c$ or  $\alpha \equiv 1/(4\pi \epsilon_0) \, e^2 / \hbar c$, where $e$, $\hbar$, $c$, and $\epsilon_0$ are the elementary charge, reduced Planck constant, speed of light in vacuum, and electric constant of free space, respectively. It is a dimensionless quantity and characterizes the strength of the electromagnetic interaction between elementary charged particles. Its possible spacetime variation can be measured from observations of distant astrophysical objects or radiation. 

Probing the $\alpha$ variation has a long history \citep[e.g.,][]{Savedoff1956,Bahcall1967}. It is now clear that any relative $\alpha$ variation $\Delta\alpha / \alpha$ in the local Universe or time variation {$\dot{\alpha} / \alpha$ ($\dot{\alpha} \equiv {\rm d}\alpha / {\rm d}t$) at the present time, if it exists, must be extremely small \citep[e.g.,][]{Damour1996,Petrov2006,Rosenband2008,Murphy2022}. Therefore, astrophysical observations mostly focus on high-redshift objects and explore the $\alpha$ variation at early times when a variable $\alpha$ is theoretically more possible \citep[e.g.,][]{Barrow2002,Alves2018}. The majority of the observational studies in the past three decades utilized quasar absorption lines based on high-resolution spectral observations. This powerful approach uses very bright quasars as background sources and analyzes absorption lines from intervening gas clouds. The absorption lines used in early works were mainly fine-structure doublet lines such as \ion{C}{4}, \ion{N}{5}, \ion{Mg}{2}, and \ion{Si}{4} \citep[e.g.,][]{Potekhin1994,Cowie1995,Murphy2001_MN327_1237,Chand2005}. In this so-call the alkali-doublet (AD) method, the wavelength separation of each doublet is directly related to $\alpha$ (particularly for these relatively light elements) and thus provides an excellent tracer of the $\alpha$ variation. 

While the AD method is relatively clean and straightforward, it does not use all the information in quasar spectra, since one absorption system usually contains a series of absorption lines. To make use of all (or most) absorption lines in an absorption system, the many-multiplet (MM) method was introduced \citep[e.g.,][]{Dzuba1999PhRvL_82_888,Webb1999}. This method takes advantage of different relativistic effects on the $\alpha$ measurement for different elements. In particular, heavy elements usually have strong relativistic effects and some of them show opposite relativistic corrections \citep{Dzuba1999PhRvA_59_230}. By comparing the absorption lines from these heavy atoms (commonly used lines include \ion{Fe}{2}, \ion{Zn}{2}, \ion{Ni}{2}, \ion{Cr}{2}, etc.) and light atoms, the MM method can achieve a much greater accuracy compared to the AD method. Therefore, later studies are mostly based on the MM method \citep[e.g.,][]{Molaro2013,Songaila2014,Wilczynska2015,Kotus2017,Milakovic2021}, and so far the strongest constraints (null results) on the $\alpha$ variation are also from this method. 

As the MM method uses all absorption lines in quasar spectra, it requires not only accurate laboratory wavelengths for all atomic transitions involved, but also accurate measurements of the line wavelengths in the observed spectra. There could be a number of systematic uncertainties, including isotope effects, slit effects, kinematic effects, velocity or wavelength distortions, etc., that may cause false detections and these systematics have been extensively discussed in the literature \citep[e.g.,][]{Murphy2001_MN_327_1223}. Some systematic uncertainties can be suppressed by large samples. For example, the absorption lines of different atoms from one absorption system may originate from different parts of a gas cloud and thus have different velocities. It is quite common in astrophysical objects and this effect can be reduced using a large number of absorption systems. One intriguing result from previous studies of quasar absorption lines was tentative evidence for both temporal and spatial variation of $\alpha$ at a level of $\Delta\alpha / \alpha \sim (1-10) \times 10^{-6}$  \citep[e.g.,][]{Murphy2001_MN_327_1208,Webb2001,Murphy2003,Webb2011,King2012}. Long-range wavelength distortions were later found to be the dominant source of systematic uncertainty in those measurements \citep[e.g.,][]{Whitmore2015}, but their overall impact may not be as severe as initially thought \citep[e.g.,][]{Dumont2017}. With the advances of precision radial velocity measurements on large telescopes such VLT ESPRESSO and new analysis methods such as artificial intelligence, the wavelength measurement is being improved and systematics associated with the wavelength calibration are being alleviated \citep[e.g.,][]{Milakovic2020,Lee2021a,Lee2021b,Schmidt2021,Milakovic2024,Schmidt2024}.

In addition to the lines in the optical, atomic fine-structure lines and molecular lines in the sub-millimeter and radio bands have also been applied to probe the $\alpha$ variation \citep[e.g.,][]{Levshakov2017,Kanekar2018}, but they have not been widely used. There are other astronomical methods, including galaxy clusters, type Ia supernovae, and cosmic microwave background \citep[e.g.,][]{deMartino2016,Holanda2016,Hart2018,Smith2019}. These methods are usually model dependent with astronomical assumptions, and have not yet achieved stringent constraints on the $\alpha$ variation. 

Despite the fact that quasar absorption lines are currently the most widely-used method, the earliest astrophysical observations actually used emission lines, particularly quasar/AGN emission lines \citep[e.g.,][]{Savedoff1956,Bahcall1967}. Emission lines in astronomy provide a great advantage compared to absorption lines: they are much easier to observe with high signal-to-noise ratios (S/Ns). The emission lines of the SDSS quasars have been used to constrain the $\alpha$ variation \citep{Bahcall2004,Albareti2015}, though  the constraint is not very stringent. The main emission line in this work is the [\ion{O}{3}] $\lambda\lambda$4959,5007 doublet (hereafter [\ion{O}{3}]). The [\ion{O}{3}] doublet is arguably the best emission line to probe the $\alpha$ variation (see Section \ref{sec:data} for a detailed explanation). The problem is that quasar emission lines are broad, and the [\ion{O}{3}] doublet lines are often affected by the H$\beta$ emission line. They also suffer from other issues, such as \ion{Fe}{2} contamination (see Section \ref{sec:data}). In emission-line galaxies (ELGs), however, the [\ion{O}{3}] doublet lines are much narrower and cleaner, and thus offer a great tool in searches for a variable $\alpha$. On the other hand, they require relatively higher spectral resolution. Because there is a lack of large samples with medium to high resolution spectra, ELGs have been rarely used so far.

We are carrying out a program to probe the $\alpha$ variation using massive galaxy spectroscopic surveys from the Dark Energy Spectroscopic Instrument \citep[DESI;][]{Levi2013,DESI2016a,DESI2016b,DESI2022_overview}. In this paper we present our first constraint using a large sample of $>100,000$ ELGs at $z\le0.95$. At $z=0.95$, the Universe was roughly half the current age. We primarily rely on the [\ion{O}{3}] doublet in this work, and will include the [\ion{Ne}{3}] $\lambda\lambda$3869,3967 (hereafter [\ion{Ne}{3}]) doublet and possibly other doublets in the next work. The layout of the paper is as follows. In Section \ref{sec:data}, we will introduce our galaxy sample and spectroscopic data. In Section \ref{sec:results}, we will measure $\Delta\alpha / \alpha$ from the galaxies and present our main results. We will discuss our results in Section \ref{sec:discussion} and summarize the paper in Section \ref{sec:summary}. Throughout the paper, all magnitudes are expressed on the AB system. We use a $\Lambda$-dominated flat cosmology with $H_0=69\,{\rm km\,s}^{-1}\,{\rm Mpc}^{-1}$, $\Omega_{\rm m}=0.3$, and $\Omega_{\Lambda}=0.7$.

\section{Data and wavelength calibration} \label{sec:data}

In this section, we will first explain the [\ion{O}{3}] doublet as an excellent probe of the $\alpha$ variation (Figure \ref{fig:oiii}). We will then introduce the DESI spectroscopic survey and our galaxy sample selection from DESI. A key step in all previous studies is the wavelength calibration. We will use sky lines to refine wavelength calibration for a small sample of our galaxies.

\subsection{[\ion{O}{3}] Doublet as a Probe of Varying $\alpha$}

\begin{figure*}[t]   
\plotone{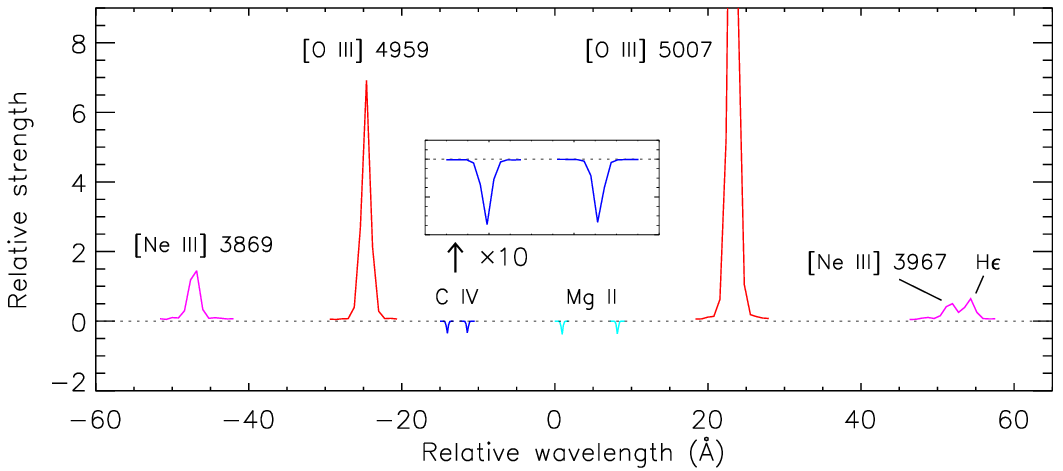}
\caption{Power of the [\ion{O}{3}] doublet by a comparison with [\ion{Ne}{3}] (magenta) and two doublet absorption lines \ion{C}{4} (blue) and \ion{Mg}{2} (cyan). The [\ion{O}{3}] and [\ion{Ne}{3}] lines are from the same low-redshift galaxy. Note that the [\ion{Ne}{3}] $\lambda$3967 line is blended with H$\epsilon$. We have assumed a resolving power of $R=20,000$ for \ion{C}{4} and \ion{Mg}{2} for the purpose of clarity. \ion{C}{4} and \ion{Mg}{2} are the most abundant (and often the strongest) metal absorption lines in quasar spectra. Compared to [\ion{O}{3}], however, they are usually weaker by orders of magnitude. \label{fig:oiii} }
\end{figure*}

As for the alkali-like doublets used in the AD method mentioned in Section \ref{sec:intro}, the wavelength separation of the [\ion{O}{3}] doublet lines $\Delta \lambda = \lambda _2- \lambda_1$ (where $\lambda_1$ and $\lambda_2$ are the line wavelengths) is directly related to $\alpha$. With a reasonable, non-relativistic approximation, $\Delta \lambda / \bar{\lambda} \propto \alpha^2$, where $\bar{\lambda}$ is the average of $\lambda_1$ and $\lambda_2$. In this work, we replace $\bar{\lambda}$ by $\lambda_2$, because the [\ion{O}{3}] $\lambda$5007 line is about three times stronger than the [\ion{O}{3}] $\lambda$4959 line and thus has a much higher S/N. This does not affect the calculation of $\Delta\alpha / \alpha$. Then $\Delta\alpha / \alpha$ is calculated by 
\begin{equation}
\Delta\alpha / \alpha \approx 0.5 \times \left( \frac{\Delta \lambda(z) / \lambda_2(z)}{\Delta \lambda(0) / \lambda_2(0)} - 1 \right),
\end{equation}
where 0 and $z$ label two redshifts. Any uniform shift of a spectrum caused by either a cosmological redshift or a Doppler effect does not influence the measurement. The present-day vacuum wavelengths of the doublet lines, typically used as the reference values, are $\lambda_1(0) = 4960.295$ \AA\ and $\lambda_2(0) = 5008.240$ \AA, respectively\footnote{https://www.nist.gov}. The uncertainties of the two values are unclear. They are barely sufficient for this work and should be improved in the future. On the other hand, the absolute values are not critical if we have a range of redshifts \citep{Bahcall2004}, because one can use Equation 1 to compare any two redshifts or two systems. 

Among all UV and optical emission lines, the [\ion{O}{3}] doublet lines are the best choice for the purpose of detecting a varying $\alpha$. The reason is twofold. The first one is that the doublet lines have a very wide wavelength separation $\Delta\lambda$ (nearly 50 \AA), roughly one order of magnitude larger than most fine-structure doublet lines. This wavelength separation is directly proportional to the sensitivity of the $\Delta\alpha / \alpha$ measurement: this is because Equation 1 can be rewritten as $\Delta\alpha / \alpha \approx 0.5\times \delta(\Delta\lambda) / \Delta\lambda$, where $\delta(\Delta\lambda)=\Delta \lambda(z)-\Delta \lambda(0)$. From this formula, a $\Delta\lambda$ change of 0.01 \AA\ for [\ion{O}{3}] means $\Delta\alpha / \alpha \approx 10^{-4}$. In other words, a systematic or measurement uncertainty of 0.01 \AA\ sets a detection limit of  $\Delta\alpha / \alpha \approx 10^{-4}$. As we will see in Section \ref{sec:results}, the accuracy of 0.01 \AA\ is roughly the best that we can achieve for individual [\ion{O}{3}] doublet systems in the DESI data.

The other reason is that the [\ion{O}{3}] doublet lines are very strong in many galaxies. In some galaxies they are the strongest emission lines in the UV and optical range, far stronger than any other doublets. Therefore, it is easy to obtain high S/N spectra for [\ion{O}{3}], which is critical for the $\Delta\alpha / \alpha$ measurement. Figure \ref{fig:oiii} demonstrates the power of the [\ion{O}{3}] doublet. In this figure, we compare a [\ion{O}{3}] doublet with [\ion{Ne}{3}] and two doublet absorption lines \ion{C}{4} and \ion{Mg}{2}. The [\ion{O}{3}] and [\ion{Ne}{3}] lines are from the same low-redshift galaxy. \ion{C}{4} and \ion{Mg}{2} are the most abundant (and often the strongest) metal absorption lines in quasar spectra. Compared to [\ion{O}{3}], however, they are usually weaker by orders of magnitude. 

As mentioned in Section \ref{sec:intro}, currently the most widely used method is based on quasar absorption lines. Emission lines have been rarely used in previous work. Quasar emission lines (mainly [\ion{O}{3}]) were used before \citep[e.g.,][]{Bahcall2004,Albareti2015}. While the [\ion{O}{3}] emission lines are usually strong in quasars, they are difficult to use for an accurate measurement of the variation in $\alpha$ due to the following reasons. The first reason is that quasar emission lines are broad and broad lines are not optimal for an accurate radial velocity measurement. Although the [\ion{O}{3}] lines are not typical broad emission lines such as \ion{C}{4} and \ion{Mg}{2} in quasars, they are generally much broader than those in galaxies. Because lines are broad, the [\ion{O}{3}] lines are often blended with the broad H$\beta$ emission line, which can affect the measurement of the line centers. Furthermore, there are plenty of \ion{Fe}{2} emission lines that form a pseudo continuum in quasar UV and optical spectra \citep[e.g.,][]{Vanden2001,Vestergaard2001,Tsuzuki2006,Jiang2007,Wang2022}. The strength of the \ion{Fe}{2} emission varies from one quasar to another. The existence of the \ion{Fe}{2} emission affects both flux and wavelength measurements for other emission lines. In quasar studies, the contaminant H$\beta$ and \ion{Fe}{2} emission can be modeled and subtracted, but the accuracy requirement is far lower than the level that we demand in this work.

\subsection{DESI Survey and Sample Selection}

ELGs are a much better probe of $\alpha$ compared to quasars, as their lines are much narrower and cleaner. The SDSS BOSS and eBOSS surveys \citep{Dawson2013,Dawson2016} produced a large number of ELGs. These galaxies were primarily used to study the expansion history of the Universe via the measurement of the baryon acoustic oscillations. The resolving power of the SDSS spectra is about 2000, barely enough to resolve individual, narrow emission lines from most galaxies. Therefore, SDSS galaxies have not been widely used to study the $\alpha$ variation. Now DESI is accumulating the largest number of galaxy spectra. Its resolving power reaches $\ge5000$ in the long wavelength range. A higher wavelength resolution means not only a better wavelength determination, but also less contamination from sky lines. 

DESI is a robotic, fiber-fed, highly multiplexed spectroscopic surveyor that operates on the Mayall 4-meter telescope at Kitt Peak National Observatory \citep{DESI2022_overview,DESI2023_EDR,DESI2023_Validation,Schlafly2023}.
The primary goal of DESI is similar to that of BOSS and eBOSS, i.e., to study the expansion history of the Universe and the growth of the large-scale structure. DESI selected targets from an imaging survey that covers more than 14,000 deg$^2$ of the sky \citep{Zou2017,Dey2019}. Its main extragalactic targets \citep{Myers2023_DESItargets} include luminous red galaxies \citep[LRGs;][]{Zhou2023_DESI_LRG}, ELGs \citep{Raichoor2023_DESI_ELG}, and quasars \citep{Alexander2023,Chaussidon2023_DESI_QSO}. These targets are used as the tracers of the large-scale structure. In addition, DESI also carries out a Bright Galaxy Survey \citep[BGS;][]{Hahn2023_DESI_BGS,Lan2023,Juneau2024} and a Milky Way Survey \citep{Cooper2023_DESI_MWS}. When we select our galaxies for this work below, we do not discriminate between different samples. We simply select galaxies with narrow and strong [\ion{O}{3}] lines from all extragalactic samples.

DESI has a large field-of-view of 8 deg$^2$, equipped with 5000 fibers for science targets in the focal plane. Light is split into three arms and recorded by three different cameras. The three arms are denoted as blue ``B'' (3600-5800 \AA), red ``R'' (5760-7620 \AA),  and near-IR ``Z'' (7520-9820 \AA), respectively. Wavelength ranges from adjacent arms slightly overlap for the purpose of calibration. The overlap regions have a much lower throughput, so we exclude galaxies whose [\ion{O}{3}] lines are located in the central parts of the overlap regions.

We select galaxies from a DESI internal value-added catalog (v2.0) generated by \texttt{FastSpecFit}\footnote{https://fastspecfit.readthedocs.org/} \citep[J. Moustakas et al. in preparation;][]{Moustakas2023}. 
We focus on strong and narrow [\ion{O}{3}] emitters, since the accuracy of the wavelength measurement is dependent of both line strength and line width. We define a line quality parameter $Q \equiv f / \sqrt{\sigma}$, where $f$ is the flux of the [\ion{O}{3}] $\lambda$4959 line in units of $10^{-17}\,{\rm erg\,s}^{-1}$ and $\sigma$ is the velocity dispersion in units of $\rm km\,s^{-1}$. If a line profile is a Gaussian, the full-width at half maximum (FWHM) of the line is $2.35\,\sigma$. Note that the [\ion{O}{3}] $\lambda$5007 line is three times stronger than the [\ion{O}{3}] $\lambda$4959 line, and they have the same intrinsic line profile. The redshift range considered in this work is $0<z<0.95$, and the upper limit of the redshifts is determined by the wavelength coverage of the Z arm. Our initial selection in this work includes galaxies with $Q>10$ and $\sigma<170\,{\rm km\,s}^{-1}$ ($\rm FWHM<400\,km\,s^{-1}$) and galaxies with $5<Q<10$ and $\sigma<130\,{\rm km\,s}^{-1}$ ($\rm FWHM<300\,km\,s^{-1}$). Our final sample contains $\sim 110,000$ galaxies after we remove a small fraction of the sources for a variety of reasons (Section \ref{sec:results}). The redshift distribution of the galaxies is shown in Figure \ref{fig:redshift}. Most of them are at relatively lower redshifts.

\begin{figure}[t]   
\includegraphics[width=0.48\textwidth]{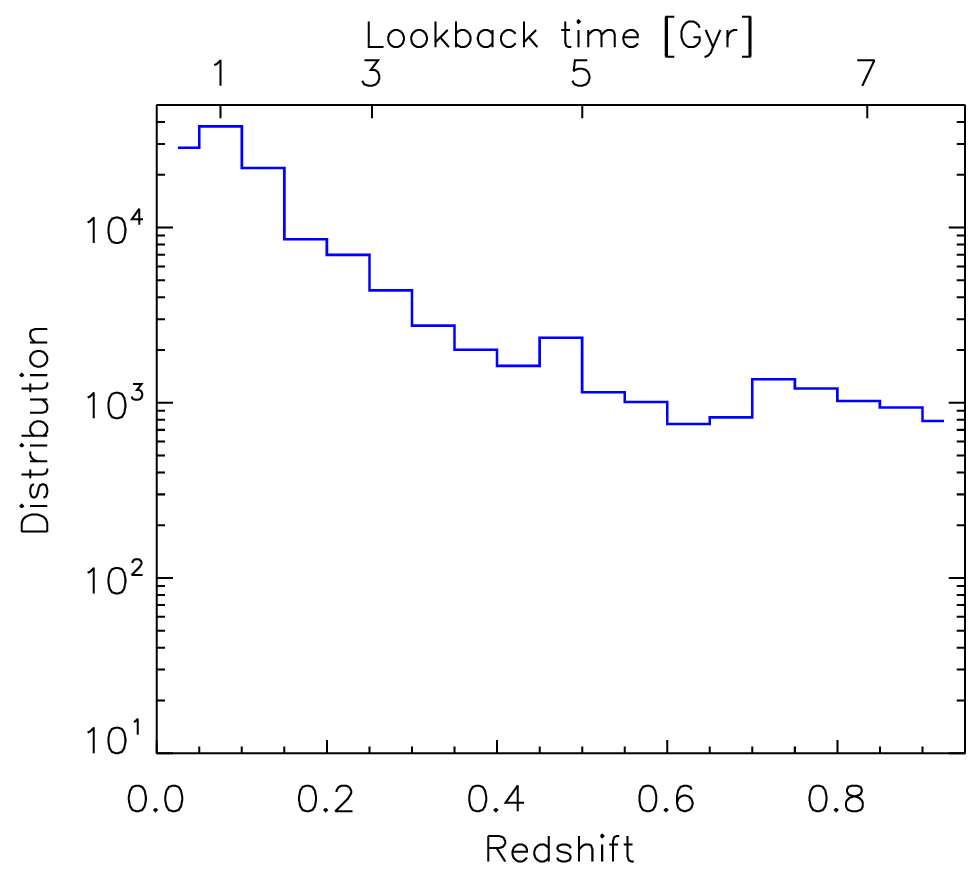}
\caption{Redshift distribution of our galaxy sample. Most galaxies are at relatively lower redshifts. The highest redshifts in the sample correspond to roughly half the cosmic time. \label{fig:redshift} }
\end{figure}

\subsection{Wavelength Calibration}

\begin{figure*}[t]   
\plotone{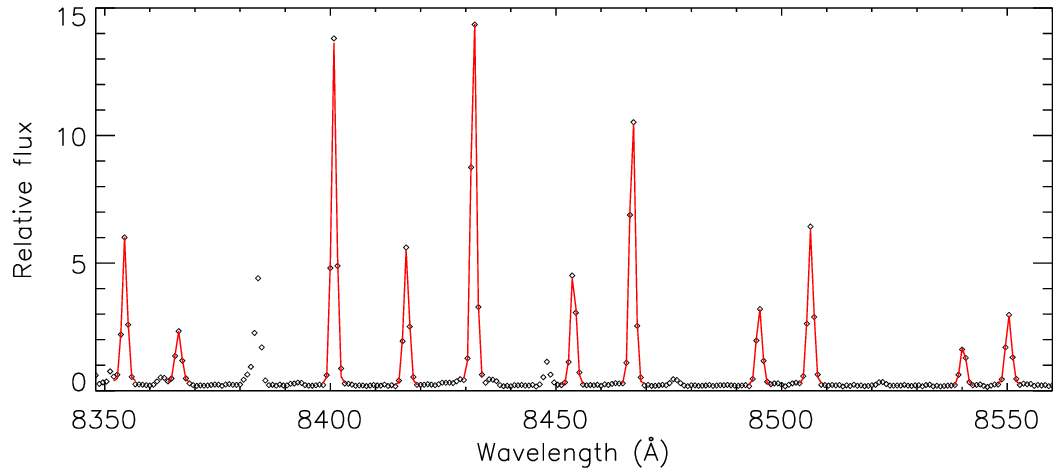}
\caption{Example of Gaussian fits to sky lines in a short wavelength range. The diamonds represent data points and the red profiles are the best Gaussian fits. A single Gaussian usually works well. Weak lines or blended lines are not used. Note that OH vibration-rotation lines are doublets owing to $\Lambda$-splitting, but the separations of the doublets are small and are not resolved in the DESI spectra.  \label{fig:skylines} }
\end{figure*}

Wavelength calibration is a critical step for the study of the variation in $\alpha$. It is particularly important for the MM method, as it deals with absorption lines in a wide wavelength range. It is often less important in the AD method, because the wavelength separations of most doublets are very small (a few \AA\ in the rest frame). However, the [\ion{O}{3}] doublet lines have a wide wavelength separation of nearly 50 \AA, which reaches nearly 100 \AA\  in the observed frame for galaxies at $z\sim1$, the highest redshift in this sample. While the wide wavelength separation greatly improves the detection sensitivity, a wavelength or velocity distortion can be obvious in such a wide wavelength range.

DESI performs its wavelength calibration in three steps, including an initial guess, a calibration with arc lamp lines, and a calibration with sky lines. It reaches a high accuracy of $\rm{rms} \sim0.02\,\AA$\ \citep{Guy2023}, which is good enough for galaxy and cosmology science programs. We test the wavelength calibration using some of the brightest [\ion{O}{3}] emitters based on the method that will be addressed in Section \ref{sec:results}. Specifically, we calculate $\Delta\alpha / \alpha$ for these galaxies. Because previous studies have demonstrated a non-variation of $\alpha$ at a level of $\sim 5\times10^{-6}$ or $\sim5$ ppm (ppm: one part per million) \citep[e.g.,][]{Lee2023,Webb2024}, we do not expect to detect a varying $\alpha$ from individual galaxies here. From the above test we find that some galaxies show a significant $\alpha$ variation, which should be caused by a wavelength distortion. We try to refine the wavelength calibration using sky lines (particularly OH lines) if an [\ion{O}{3}] doublet is located in a wavelength range with sufficient sky lines. We do not try to derive a new global wavelength solution for each galaxy. Instead, we calculate the wavelengths of the sky lines and estimate their deviations from their theoretical or lab values. 

For each galaxy with an [\ion{O}{3}] doublet surrounded by sky lines, we start with its two intermediate files ``sframe-CAMERA-EXPID.fits" file (``sframe" file for short) and ``sky-CAMERA-EXPID.fits" file (``sky" file for short), where CAMERA is one of the spectrograph cameras and EXPID is a 8-digit exposure ID. The ``sky" file contains the sky model for the galaxy and the ``sframe" file contains the sky-subtracted galaxy spectrum before the step of the flux calibration. We combine the ``sframe" and ``sky" files to produce a galaxy spectrum with sky lines. We then select isolated and strong sky lines in the spectrum to calculate their wavelengths. A Gaussian profile is fitted to each line. A single Gaussian usually works well. Figure \ref{fig:skylines} shows an example. Sky lines are mostly made of OH vibration-rotation lines and they are doublets owing to $\Lambda$-splitting with the same flux. The separations of the doublets are typical a few tenths of  \AA, so either individual lines or doublets are not resolved by DESI. Therefore, a more complex sky line model is not demanded. 

We perform a series of tests regarding how to select sky lines. First of all, we only consider strong and isolated lines. Weak lines produce large uncertainties and do not help us improve the wavelength calibration. Second, some galaxies were observed in bright nights and/or with short exposure times (much shorter than the nominal time of 1000 seconds). In this case, some strong sky lines appear weak in the spectra and are excluded in our analyses. Finally, we note that we can directly fit the lines in the sky model spectrum rather than the combined spectrum. We do a test and find that the results are well consistent. Since the sky model spectrum usually has a higher S/N, we decide to use the sky model spectrum to derive the sky line wavelengths and the whole procedure is the same.

After we obtain the wavelengths of the sky lines, we compare them with lab wavelengths. \citet{Abrams1994} measured OH line wavelengths between 1850 and $\rm 9000\,cm^{-1}$ in the lab to a great accuracy (better than $\rm 0.00001\,\AA$), and their result is one of the calibration standards. Their measurements were done under conditions similar to those in the upper atmosphere where OH lines are produced. We do not directly use the result of \citet{Abrams1994}, because most of their measured lines are very weak (or not detected) in the DESI spectra. Instead, we pick up strong lines from the list of \citet{Rousselot2000}. \citet{Rousselot2000} measured the wavelengths of OH lines using spectra taken by the VLT telescope, and their wavelengths were calibrated to match those in \citet{Abrams1994}. In addition to the OH lines, there are some other lines (such as O$_2$ and \ion{O}{1}) in the wavelength range that we consider. These lines were not included in the above sky line list. We add several of these isolated and strong lines using the result of \citet{Osterbrock1996,Osterbrock1997}. \citet{Osterbrock1996} measured sky lines using spectra taken by the Keck telescope, and their wavelength calibration was also done by matching the result of \citet{Abrams1994}. Note that DESI uses the wavelength measurements of \citet{Hanuschik2003} in the last step of its wavelength calibration. The result of \citet{Hanuschik2003} is consistent with previous studies mentioned above. We also use \citet{Hanuschik2003} to crosscheck our result. Section \ref{subsec:waveRefine} will describe how the new wavelengths are applied to our analyses.

It is worth mentioning that wavelengths values are either air or vacuum wavelengths in the literature. Different works used different conversion formulas between air and vacuum wavelengths \citep[e.g.][]{Edlen1953,Edlen1966,Ciddor1996}. For example, \citet{Abrams1994} claimed that they used the \citet{Edlen1966} conversion. Later \citet{Osterbrock1996} found that \citet{Abrams1994} actually used the \citet{Edlen1953} conversion. \citet{Rousselot2000} stated that they used the \citet{Allen1973} conversion, which is actually based on the \citet{Edlen1953} conversion. \citet{Hanuschik2003} used the \citet{Edlen1966} conversion. We use a relatively new conversion by \citet{Ciddor1996}. Nevertheless, the differences among these conversions are tiny; see \citet{Murphy2001_MN_327_1208} for a detailed comparison. These differences can be completely ignored in this work.

\section{Results} \label{sec:results}

\begin{figure*}[t]   
\plotone{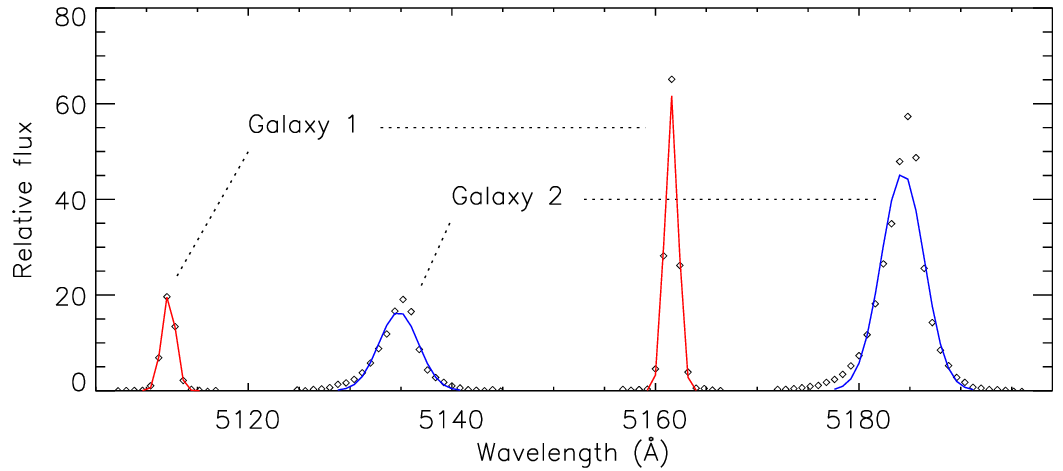}
\caption{Example of a single Gaussian fit to [\ion{O}{3}]  in two galaxies, including a good fit to a narrow [\ion{O}{3}] doublet and a poor fit to a broader [\ion{O}{3}] doublet (due to its asymmetric line profile). The diamonds represent the data points and the underlying continuum emission has been subtracted. The red and blue profiles are the best Gaussian fits.  \label{fig:1gfit} }
\end{figure*}

In this section, we measure $\Delta\alpha / \alpha$ values for our galaxy sample and present our main results. We use Equation 1 to calculate $\Delta\alpha / \alpha$, meaning that we need to measure the separation of the [\ion{O}{3}] doublet lines $\Delta\lambda(z)$ and the wavelength of the [\ion{O}{3}] $\lambda$5007 line $\lambda_2(z)$. The two doublet emission lines intrinsically have the identical line profiles as both transitions are from the same excited energy level of double-ionized oxygen. However, this does not mean that the two lines have the same profiles in an observed spectrum, because the spectral resolution and dispersion could be slightly different at two different wavelengths. 
In addition, a doublet line could be the superposition of multiple components, as we will see in Section \ref{subsec:gfit}. In this case, the two line profiles are still intrinsically identical, but the wavelength measurement becomes complex. A direct cross correlation is not necessarily a good method to find the separation of the doublet lines. \citet{Bahcall2004} used a modified cross-correlation method, i.e., the relative strength and width of the two lines were allowed to vary. This method worked for quasar emission lines that are broad and cover many pixels. It does not work well for our narrow emission lines that cover only several pixels (see Galaxy 1 in Figures \ref{fig:1gfit} and \ref{fig:2gfit}), because the fine resampling algorithm for narrow lines during the cross correlation strongly relies on the assumption of intrinsic line profiles. Therefore, we prefer direct model fitting to the lines.

We will first use a single Gaussian to fit individual [\ion{O}{3}] doublet lines and then use double Gaussian profiles to improve the line fit. We will finally use sky lines to refine our measurements when possible. In this work we will focus on the time variation of $\alpha$, but will also briefly estimate its spatial variation.

\subsection{Single Gaussian Fit and Double Gaussian Fit} \label{subsec:gfit}

\begin{figure*}[t]   
\plotone{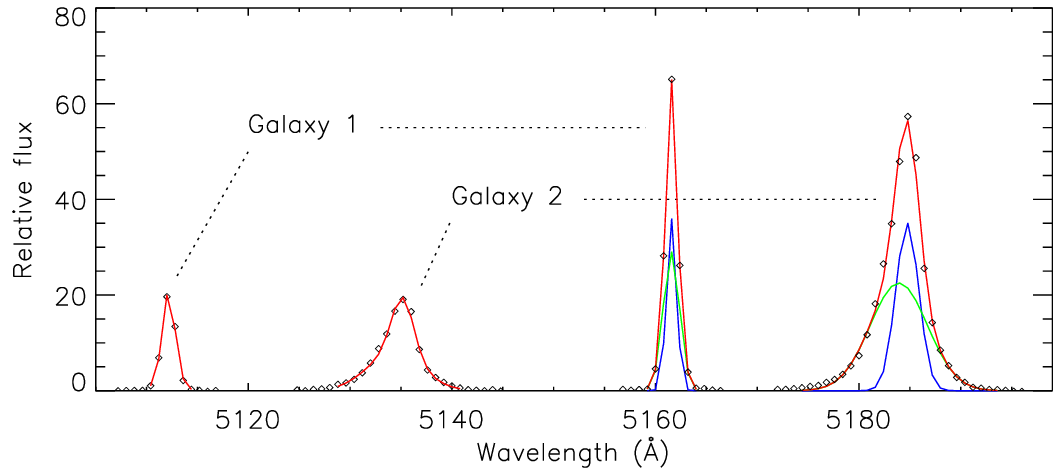}
\caption{Example of a double Gaussian fit to [\ion{O}{3}]  in two galaxies. The two galaxies are the same as shown in Figure \ref{fig:1gfit}. The diamonds represent the data points and the underlying continuum emission has been subtracted. The red profiles are the best Gaussian fits. For the two [\ion{O}{3}] $\lambda$5007 lines, we also show individual Gaussian components in green and blue. A double Gaussian usually provides a good fit to our sample galaxies. \label{fig:2gfit} }
\end{figure*}

For each emission line, we first subtract its underlying continuum emission, which is obtained by fitting a second-order polynomial curve to a short spectral range around the line. In this work we only selected strong [\ion{O}{3}] emitters described in the previous section, and the majority of the galaxies in our sample have relatively weak continuum emission and thus large equivalent widths (EWs). After the continuum subtraction, we fit a Gaussian to the line. The two [\ion{O}{3}] doublet lines are fitted independently. Figure \ref{fig:1gfit} shows an example of our best single Gaussian fits to two galaxies, including a good fit to a narrow [\ion{O}{3}] doublet and a relatively poor fit to a broader [\ion{O}{3}] doublet. The catalog that we use contains a small number of contaminants. In addition, some galaxies or emission lines were misclassified. We remove them in this step. 

A single Gaussian works reasonably well for narrow [\ion{O}{3}] emitters. We also try to fit a Voigt profile that has one more free parameter, and the results are generally poorer. As the line width increases, a single Gaussian works less well, mainly because the lines show more asymmetric profiles, i.e., a line is presumably a superposition of more than one component under the DESI spectral resolution. Galaxy 2 in Figure \ref{fig:1gfit} clearly shows an asymmetric profile. We will see in Figure \ref{fig:2gfit} that this galaxy can be well fitted by two Gaussian components. 

In order to improve the performance of the line fitting, we add one more Gaussian component in the line model. For each emission line, we  subtract its underlying continuum emission as we did above. We first perform a double Gaussian fit to the [\ion{O}{3}] $\lambda$5007 line, since it is three times stronger than the other doublet line. A double Gaussian fit requires at least 6 effective data points. As we can see from Figures \ref{fig:1gfit} and \ref{fig:2gfit}, narrow lines in the DESI spectra barely meet this requirement. After we obtain the best fit for the [\ion{O}{3}] $\lambda$5007 line, we fit the [\ion{O}{3}] $\lambda$4959 line using the [\ion{O}{3}] $\lambda$5007 result as a prior. During the fit, we fix the separation, the relative strength, and the relative line width of the two Gaussian components of the [\ion{O}{3}] $\lambda$4959 line as determined from the [\ion{O}{3}] $\lambda$5007 line, given that the two doublet lines are intrinsically identical. Therefore, the number of free parameters in this fit is only 3. Figure \ref{fig:2gfit} illustrates our double Gaussian fit to the same two galaxies displayed in Figure \ref{fig:1gfit}. The fitting result is significantly improved for broad lines. 

We carefully evaluate the single and double Gaussian fitting results. For each galaxy, we prefer its single Gaussian result unless the single Gaussian fit is poor as judged by its $\chi^2$ value. As expected, a single Gaussian profile works well for narrow lines (roughly $\rm FWHM \le130\, km\, s^{-1}$) and a double Gaussian model is not needed in most cases. For broader lines, a double Gaussian model works better. For the broadest lines in our sample with $\rm FWHM>250\,km\,s^{-1}$, a double Gaussian is often far better than a single Gaussian. As a result, we usually adopt the single Gaussian results for narrow lines and the double Gaussian results for broad lines.

A small fraction of the galaxies are poorly fitted by either a single Gaussian or double Gaussian. We visually inspect many of these galaxies and find that most of them have very broad and complex line profiles that cannot be described by a double Gaussian. The remaining of these galaxies include different types of objects. Here are two typical examples. The first example is that the $\lambda$4959 and $\lambda$5007 lines have apparently different line profiles. In the second example, the $\lambda$4959 line shows strong and abnormal emission or absorption features due to unknown reasons. We remove these poorly fitted lines/galaxies. For very broad and complex lines, one may in principle add one or more Gaussian components to achieve a better fit. We do not do it because the resultant uncertainties would be much larger. In addition, the overall fraction of these galaxies is small in our sample, so we just exclude them.

\subsection{Wavelength Refinement} \label{subsec:waveRefine}

We refine the measured wavelengths of the [\ion{O}{3}] line centers using sky lines. This can only be done for a small fraction of galaxies or lines with plenty of nearby sky lines in the ``R'' and ``Z'' arms (corresponding to $z\ge0.2$). This cannot be done for [\ion{O}{3}] lines in the ``B'' arm since there are no bright sky lines. The details how we calculated and calibrated the wavelengths of sky lines have been elaborated in Section 2.3. We do not try to find a new global wavelength solution for each galaxy, because the DESI pipeline has already done this. Instead, we refine the wavelength calibration for each pair of the [\ion{O}{3}] doublet lines with four or more well measured nearby sky lines. We first compute the wavelength differences between the measured values and reference values for the sky lines, as stated in Section 2.3. We then fit a second-order polynomial curve to the wavelength differences as a function of wavelength. From the best-fit curve, we estimate the wavelength differences for the [\ion{O}{3}] doublet lines and apply the differences to the wavelengths of the doublet lines.

Figure \ref{fig:waveCal} compares the results for four galaxy samples before and after the wavelength refinement, by illustrating their distributions of the relative $\alpha$ variation $\Delta\alpha / \alpha$ measurements. The four redshifts used in the figure correspond to four wavelength ranges (for [\ion{O}{3}]) with a few or more usable sky lines. The blue and red histograms represent the $\Delta\alpha / \alpha$ distributions before and after the wavelength refinement. The figure shows that the wavelength refinement only moderately improves the results, and the difference between the distributions is small. This is because only a fraction of galaxies in each sample can be improved in this step. We will further discuss the wavelength calibration and refinement in the Section \ref{sec:discussion}. 

\begin{figure}[t] 
\includegraphics[width=0.48\textwidth]{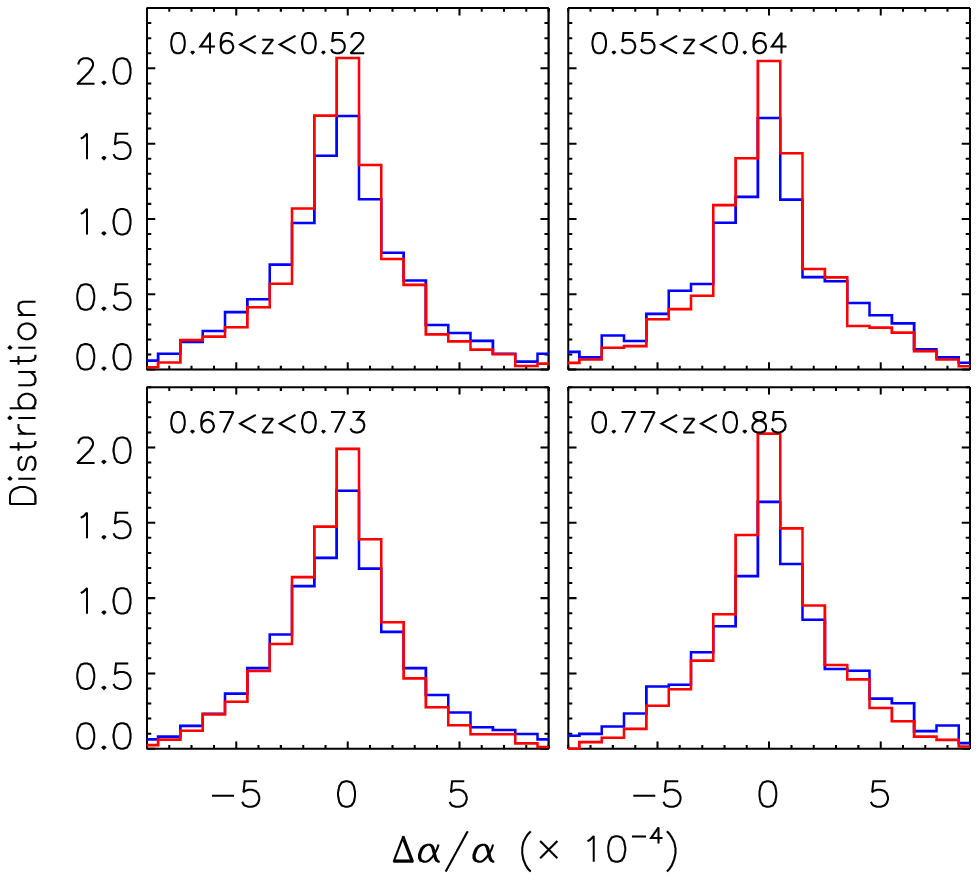}
\caption{Examples of $\Delta\alpha / \alpha$ distributions before and after the wavelength refinement (blue and red histograms, respectively). The four redshifts correspond to four wavelength ranges (for [\ion{O}{3}]) with plenty of sky lines. Each histogram has been normalized so that the total number is 10. The wavelength refinement moderately improves the results, because only a fraction of galaxies  can be improved in this step.  \label{fig:waveCal} }
\end{figure}

Figure \ref{fig:alphaDistribution} shows the $\Delta\alpha / \alpha$ distributions for the whole galaxy sample in four redshift bins. We have removed a tiny fraction of galaxies with $\Delta\alpha / \alpha>10^{-3}$. We visually inspect these galaxies and their lines suffer from serious issues mentioned in the above section. With a close inspection, the distributions in Figure \ref{fig:alphaDistribution} are not symmetric. They are slightly skewed with higher numbers at lower values. This is particularly obvious in the first panel for the sample at $z<0.2$. The asymmetric distributions naturally lead to non-zero detections of $\Delta\alpha / \alpha$, as we will see in Section 3.4. 

\begin{figure}[t] 
\includegraphics[width=0.48\textwidth]{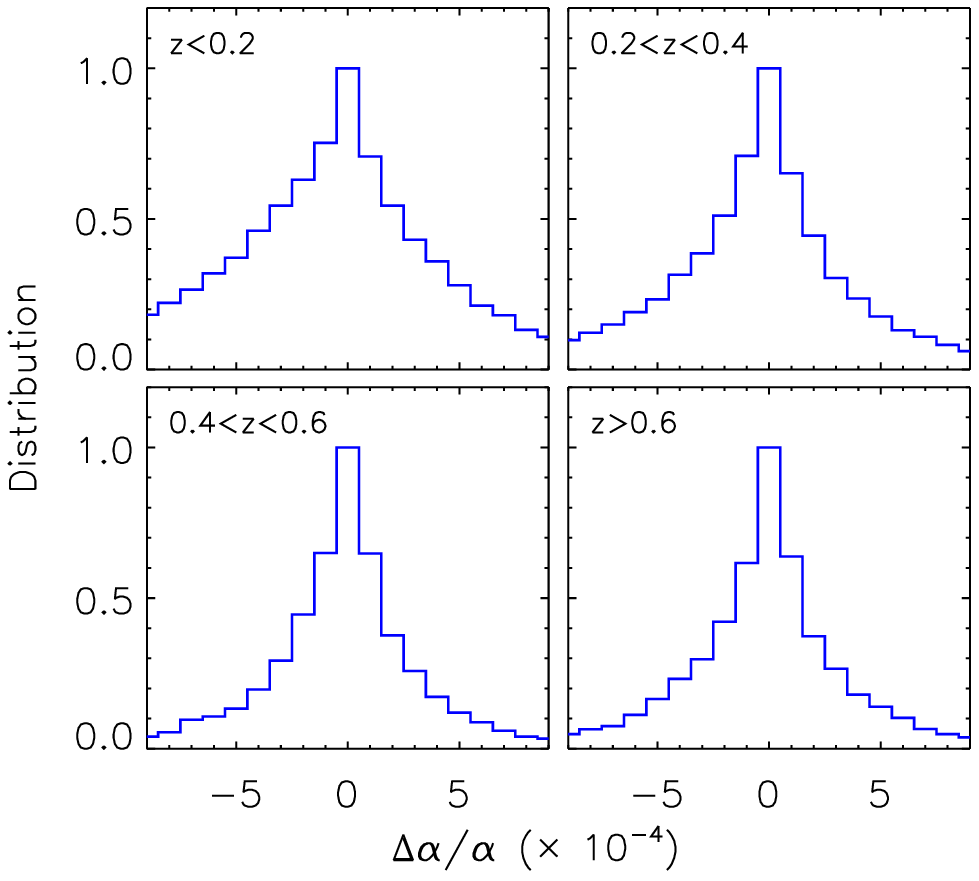}
\caption{Distributions of $\Delta\alpha / \alpha$ for the whole galaxy sample in four redshift ranges. Each histogram has been normalized so that the peak value is 1. The distributions are slightly skewed with higher numbers at lower values, which leads to non-zero detections of $\Delta\alpha / \alpha$.  \label{fig:alphaDistribution} }
\end{figure}

\subsection{Error Analyses} \label{subsec:error}

For each galaxy, the uncertainty in the $\Delta\alpha/\alpha$ measurement is the combination in quadrature of a measurement error and a systematic error from the wavelength calibration. The measurement error is estimated by Equation 1 from the propagation of the errors of the two line centers. It is dominated by the uncertainty of the $\lambda$4959 line, which is usually smaller than $\rm 0.05\,\AA$. In a tiny fraction of galaxies that have narrow lines with very high S/Ns, the measurement uncertainties of the line centers can be  smaller than $\rm 0.01\,\AA$. Most of these galaxies are at $z<0.1$ from the BGS sample. In this case, the $\Delta\alpha/\alpha$ measurement is dominated by the systematic error. 

It is not straightforward to estimate the systematic error. \citet{Guy2023} claimed that the wavelength calibration of DESI reaches an accuracy of $\rm{rms} \sim0.02$ \AA. As we will see in Section \ref{sec:discussion}, the absolute calibration is actually worse than $\rm rms \sim 0.02\,\AA$. This has a small impact on our $\Delta\alpha/\alpha$ measurement since $\Delta\alpha/\alpha$ is much more sensitive to the relative calibration in the wavelength range between the two doublet lines. However, the relative wavelength calibration is complex. In fact, this was one of the major challenges in the past. We estimate the relative wavelength calibration using the wavelength range with plenty of sky lines, and find $\rm rms \sim 0.02-0.03\,\AA$. We suspect that the accuracy of the wavelength calibration mentioned by \citet{Guy2023} is actually for the relative wavelength calibration. Therefore, we apply a uniform systematic error of $\rm 0.02\,\AA$ in our wavelength calibration. 

In the next two subsections, we will estimate the $\alpha$ variation in space and time. We divide the whole sample into different subsamples and measure $\Delta\alpha/\alpha$ for individual subsamples. For each subsample, we calculate its $\Delta\alpha/\alpha$ and related error using two methods. In the first method, we take a weighted average. This is a straightforward approach, assuming that individual input errors are reliable. The second method is the bootstrap estimate \citep[e.g.,][]{Bahcall2004}. For a subsample with $n$ galaxies, its $\Delta\alpha/\alpha$ and error are estimated from 10,000 simulated samples. To generate a simulated sample, we randomly draw $n$ galaxies with replacement from the real subsample, and calculate the weighted average of $\Delta\alpha/\alpha$ for the simulated sample. We obtain 10,000 weighted averages from the 10,000 simulated samples and they obey a Gaussian distribution. Finally, the $\Delta\alpha/\alpha$ and error values of the subsample are the mean and standard deviation of this Gaussian distribution. We will see in the next subsection that the results from the two methods are well consistent.

\subsection{Time Variation of $\alpha$} \label{subsec:timeVary}

We estimate $\Delta\alpha/\alpha$ (relative to the $z=0$ value) at different redshifts. The whole galaxy sample is divided into 10 redshift bins from $z=0$ to 1.0, with a bin size of 0.1. The $\Delta\alpha/\alpha$ values are calculated using the two methods above. The color-coded symbols in the upper panel of Figure \ref{fig:timeVary} exhibit the results of the whole sample from the two methods and they are well consistent, except that the uncertainties from the bootstrap estimate are slightly smaller. The sizes of the subsamples decrease towards higher redshifts as seen from Figure \ref{fig:redshift}, so the $\Delta\alpha/\alpha$ uncertainties increase towards higher redshifts. The uncertainties are roughly between $2\times10^{-6}$ and $2\times 10^{-5}$. Since the two methods provide nearly the same results, we will only use the weighted average method in the following analyses. In the lower panel of Figure \ref{fig:timeVary}, we divide the whole sample into two subsamples with relatively high ($Q>10$) and low ($Q<10$) line quality parameters. The results are indicated in green and magenta and they are also consistent. 

\begin{figure}[t]   
\includegraphics[width=0.45\textwidth]{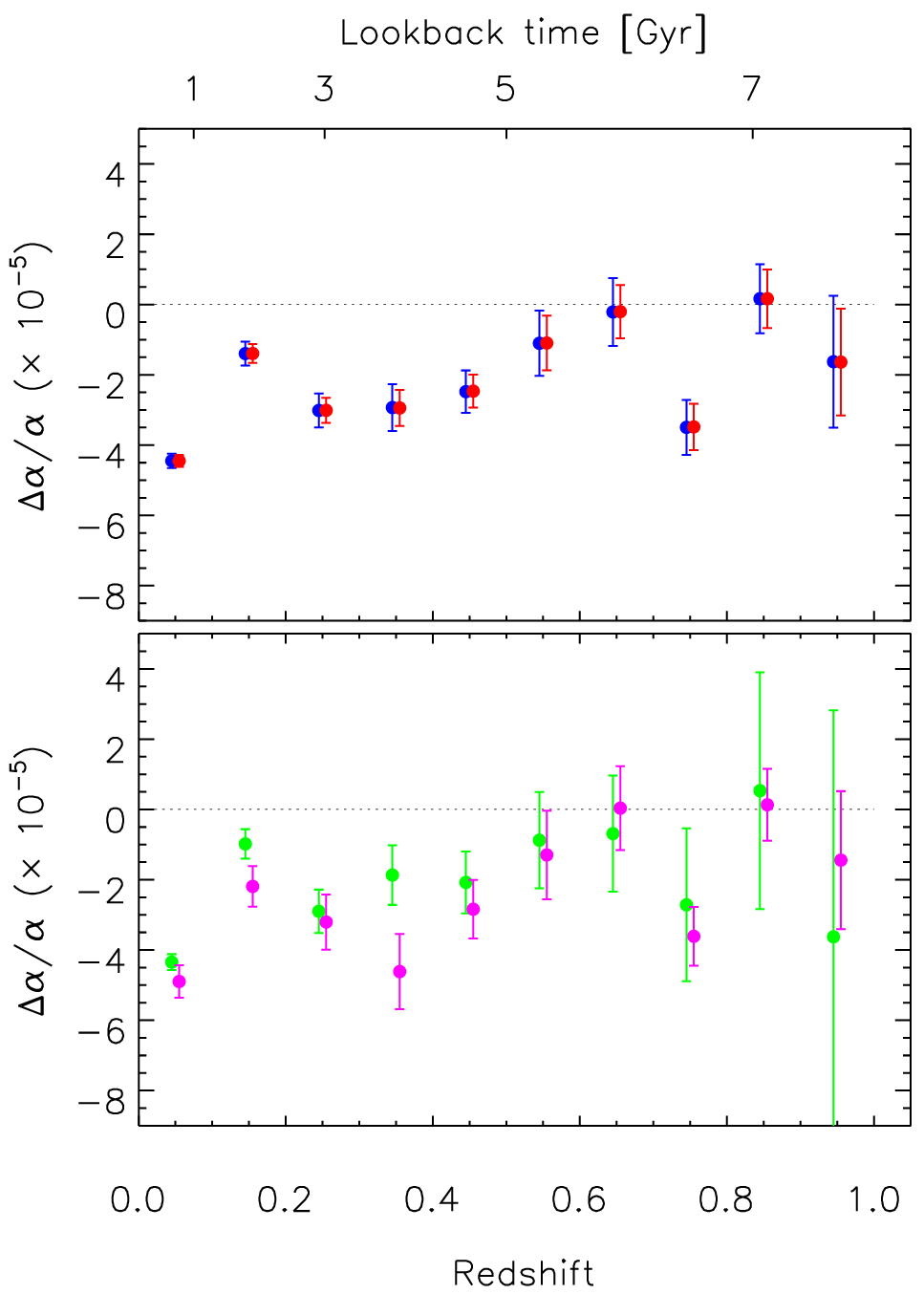}
\caption{Relative $\alpha$ variation $\Delta\alpha/\alpha$ with redshift. The blue and red circles in the upper panel represent the results of the whole sample from the weighted average and bootstrap estimate methods, respectively. The circles are slightly shifted along the $x$-axis for clarity. The two results are well consistent. In the lower panel, we show the results for two subsamples with $Q>10$ (green circles) and $Q<10$ (magenta circles), respectively, and they are also consistent. The figure shows an apparent $\alpha$ variation with time. This should be caused by systematics associated with the wavelength measurement (see Section 3.4). Note that the zero point is only relative.  \label{fig:timeVary} }
\end{figure}

Figure \ref{fig:timeVary} shows an apparent variation of $\alpha$ in most redshift bins. Specifically,  $\Delta\alpha/\alpha$ is below zero (at $\ge2\sigma$ level) at $z<0.5$ and $z\sim0.75$. As mentioned in Section 1, previous studies based on observations of bright quasars have indicated $\Delta\alpha/\alpha<10$ ppm, so the $\alpha$ variation seen in Figure \ref{fig:timeVary} should be caused by systematics associated with the wavelength measurement. If this is caused by the wavelength distortion between the two doublet lines, we can estimate the level of the distortion as follows. From Figure \ref{fig:timeVary}, the variation of $\Delta\alpha/\alpha$ is roughly $(2-3)\times10^{-5}$. Based on our estimate in Section 2.1 that an uncertainty of $\rm 0.01\,\AA$ sets a detection limit of $\Delta\alpha / \alpha \approx 10^{-4}$, $\Delta\alpha / \alpha = (2-3)\times 10^{-5}$ suggests a wavelength distortion of $\rm 0.002-0.003\,\AA$. This is beyond the detection capability of the DESI wavelength calibration. Therefore, we have reached a regime in which the constraint on the $\alpha$ variation is completely dominated by systematics. If there were no wavelength-related systematics, we would be able to achieve an accuracy of several ppm on the $\Delta\alpha/\alpha$ constraint in most redshift bins, and a stronger constraint can be obtained when all data are combined.

It is worth noting that the zero point of $\Delta\alpha/\alpha$ is only relative. It is relative to the $z=0$ value, but the uncertainty of the $z=0$ value is unclear. As  mentioned in Section 2.1, the wavelength measurement of the [\ion{O}{3}] doublet lines at $z=0$ is accurate to a few times $\rm 0.001\,\AA$. This translates to an uncertainty of a few times $10^{-5}$ for $\Delta\alpha / \alpha$, which is close to the uncertainties shown in Figure \ref{fig:timeVary}. If the $z=0$ value is very accurate, $\Delta\alpha/\alpha$ is consistent with zero at $z\sim0.65$ and $z\approx0.85$. We notice that these two redshift ranges for the [\ion{O}{3}] doublets correspond to the wavelength ranges with plenty of sky lines, which makes a robust wavelength calibration. Nevertheless, we focus on the relative $\alpha$ variation in this work.

\subsection{Spatial Variation of $\alpha$} \label{subsec:spaceVary}

Our large galaxy sample provides a great advantage that allows us to measure the spatial variation of the fine-structure constant. The current DESI data already cover a large portion of the sky. In order to estimate the spatial variation, we divide the whole sample into smaller subsamples based on redshift. The reason is twofold. The first reason is to estimate the spatial variation at different redshifts. The second reason is that this can partially remove the effect of the wavelength calibration that probably causes the time variation seen in Figure \ref{fig:timeVary}, if the wavelength calibration has a strong wavelength dependence. To efficiently remove the potential wavelength dependence, we first estimate $\Delta\alpha/\alpha$ in narrow redshift slices (narrower than those shown in Figure \ref{fig:timeVary}), and find relatively larger redshift ranges in which the measured $\Delta\alpha/\alpha$ values do not vary significantly. We then build subsamples for these large redshift ranges. Figure \ref{fig:spaceVary} shows our $\Delta\alpha/\alpha$ measurements in 6 redshift slices.

For a subsample in a redshift slice, we calculate $\Delta\alpha/\alpha$ in a grid of coordinates. The grid is made as follows. The declination range of $\rm -15\degr <Decl. < 75\degr$ is divided into 6 bins with a bin size of $15\degr$. The R.A. range is also divided into bins with different bin sizes depending on declination. From Decl. $=-15\degr$ to $75\degr$, the bin sizes are $12\degr$, $12\degr$, $15\degr$, $18\degr$, $20\degr$, and $30\degr$, respectively. This ensures that different grid cells have similar area. We require that each cell has at least 15 galaxies, otherwise this cell is not used. The central position of a cell in Figure \ref{fig:spaceVary} is the median position of the galaxies in this cell. The color-coded squares represent our $\Delta\alpha/\alpha$ measurements. The $\Delta\alpha/\alpha$ values are relative in each redshift range (or in each panel). We can see that the $\Delta\alpha/\alpha$ distribution in each panel is quite random, with a typical amplitude smaller than $10^{-4}$. The rms values of the $\Delta\alpha/\alpha$ distributions from the top to the bottom panels are 0.44, 0.29, 0.51, 0.74, 0.70, and $0.62\times10^{-4}$, respectively. The difference is mainly caused by the different subsample sizes. We do not see obvious, large-scale structures of $\Delta\alpha/\alpha$. 

\begin{figure}[t]   
\includegraphics[width=0.45\textwidth]{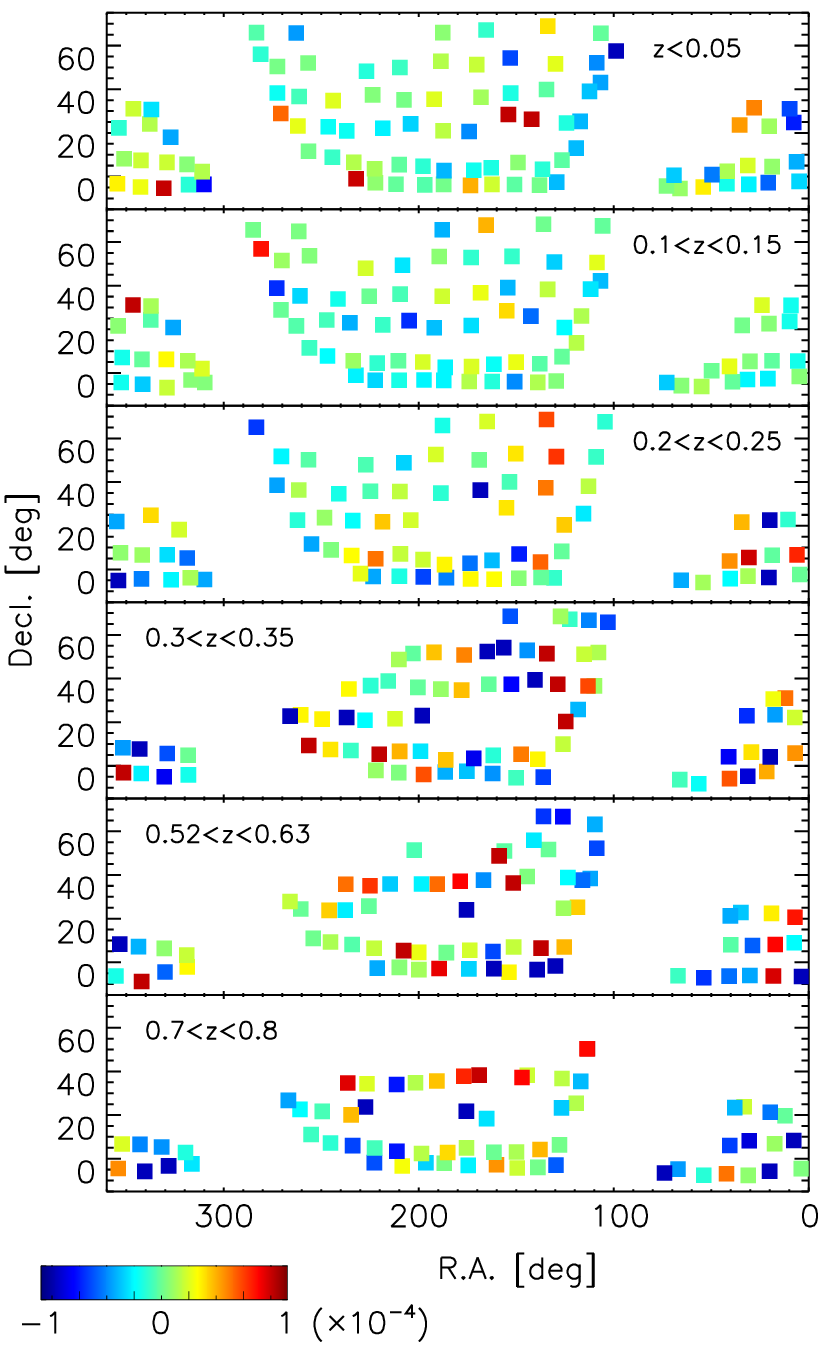}
\caption{Spatial variation of $\Delta\alpha/\alpha$. We show our $\Delta\alpha/\alpha$ measurements for 6 redshift slices in the 6 panels. In each panel, $\Delta\alpha/\alpha$ is calculated in a grid of coordinates. The color-coded squares indicate the positions of the grid cells. Each cell contains at least 15 galaxies (otherwise this cell is not used), and its position is the median position of the galaxies in this cell. The figure does not show an obvious, large-scale variation of $\alpha$. \label{fig:spaceVary} }
\end{figure}

\section{Discussion} \label{sec:discussion}

\subsection{Variation of $\alpha$ and Wavelength Calibration}

\begin{figure}[t]   
\includegraphics[width=0.45\textwidth]{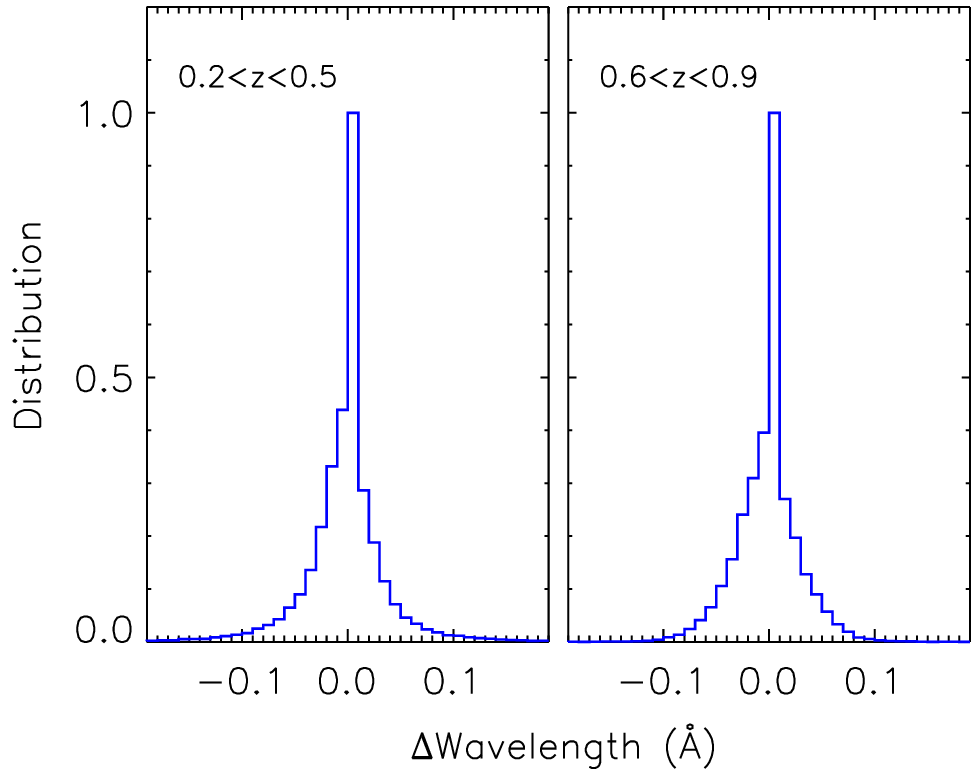}
\caption{Relative wavelength calibration in two redshift ranges. The wavelength differences shown in the $x$-axis are the differences between the measured wavelengths and reference values of sky lines in small wavelength ranges ($50-100$ \AA). Each wavelength range covers the [\ion{O}{3}] doublet lines of one galaxy. A median difference value for each galaxy has been subtracted from each range. The distributions indicate that the relative calibration is generally good, with a rms of $0.02\sim0.03$ \AA. \label{fig:testSky} }
\end{figure}

Figure \ref{fig:timeVary} in Section \ref{subsec:timeVary} shows an apparent variation of $\alpha$ with redshift. As we have briefly discussed, this is mostly likely caused by systematics associated with the wavelength calibration. We have two reasons for this claim. The first one is that the $\alpha$ variation seen in the figure can be caused by a very small wavelength distortion of $\rm 0.002-0.003\,\AA$, and the DESI wavelength calibration cannot reach such an accuracy. The second reason is that previous studies have put a strong upper limit on $\Delta\alpha/\alpha$ and the limit is roughly 5 ppm. These previous constraints were mostly from the measurements of quasar absorption lines that are produced by intervening gas. The [\ion{O}{3}] emission lines in this work are produced by the interstellar medium in star-forming galaxies. Although the absorption and emission lines are from different environments, they should obey the same laws of physics when they are at the same redshift. Because of the systematics, we did not measure the time variation of {$\dot{\alpha} / \alpha$ in Section \ref{subsec:timeVary}.

From the wavelength refinement procedure in Section \ref{subsec:waveRefine}, we find that the absolute wavelength calibration, as described by the wavelength differences between the measured values and reference values of sky lines, is poorer than what we expected. The wavelength differences sometimes reach a few tenths of \AA. The $\Delta\alpha/\alpha$ measurement is not sensitive to the absolute wavelength calibration. Instead, it mainly relies on the relative calibration. In order to quantify the relative calibration, we first calculate the above wavelength differences in small wavelength ranges ($\rm 50-100\,\AA$). Each wavelength range covers the [\ion{O}{3}] doublet lines of one galaxy. The wavelength differences in this range are subtracted by their median value, i.e., the absolute wavelength deviation is removed. Figure \ref{fig:testSky} shows the distributions of these wavelength differences in two redshift ranges. We can see that the relative calibration is generally good, with a rms of $\rm 0.02\sim0.03\,\AA$.

In Section \ref{subsec:gfit} we mentioned that narrower lines usually provide better wavelength measurements and thus better constraints on the $\alpha$ variation. We further look into the effect of the line widths on our results. The galaxy sample is divided into subsamples with $\rm FWHM<150\,km\,s^{-1}$ and $\rm FWHM>150\,km\,s^{-1}$, respectively. The upper panel of Figure \ref{fig:testWidth} shows the $\Delta\alpha/\alpha$ distributions of the two subsamples in two redshift ranges. It is clear that narrow lines (in red) provide slightly better results. The lower panel of Figure \ref{fig:testWidth} shows the redshift evolution of the $\Delta\alpha/\alpha$ measurements for the two subsamples. In the first redshift bin, we only show the narrow-line sample because the size of the other sample is too small. The results for the two subsamples are well consistent, suggesting that the effect of the line widths on our results is minor.

\begin{figure}[t]   
\includegraphics[width=0.45\textwidth]{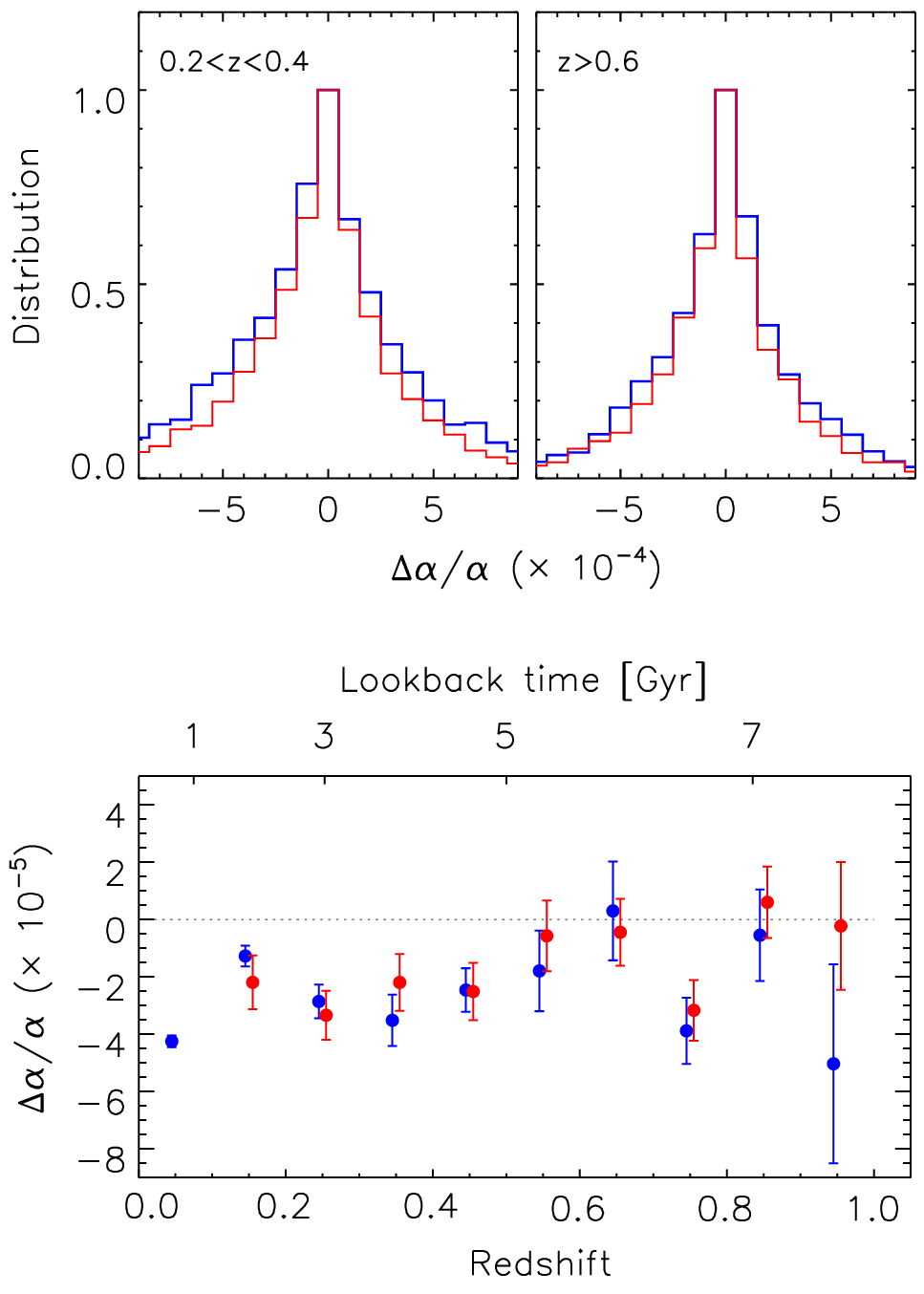}
\caption{Effect of line widths on the $\Delta\alpha/\alpha$ measurements. The upper panel shows the $\Delta\alpha/\alpha$ distributions in two redshift ranges for two galaxy samples with FWHM $<150$ km s$^{-1}$ (in red) and FWHM $>150$ km s$^{-1}$ (in blue), respectively. The lower panel shows the redshift evolution of the $\Delta\alpha/\alpha$ measurements for the two samples. \label{fig:testWidth} }
\end{figure}

\subsection{Future Prospects}

We have used about 110,000 strong [\ion{O}{3}]-emitting galaxies to constrain the $\alpha$ variation in Section \ref{sec:results}. When the galaxies were grouped into 10 redshift bins, the statistical uncertainties in the $\Delta\alpha/\alpha$ measurements are between $2\times10^{-6}$ and $2\times 10^{-5}$. The uncertainty in the lowest-redshift bin reaches $2\times10^{-6}$, owing to the largest number of galaxies and the highest average S/N in this bin. The uncertainties in other redshift ranges are mostly around $(0.5-1.0)\times 10^{-5}$. On the other hand, we have reached the regime in which our results are actually limited by the systematics related with the wavelength calibration. 

There are more systematics that may have affected our results, but their impact should be much weaker compared to those related with the wavelength calibration. Among them, the most important one is probably from line shapes. As  mentioned earlier, the two doublet lines intrinsically have the identical line profiles. For example, if one line shows an asymmetric feature caused by inflows, outflows, or other reasons, the other line will have the same feature. In the observed spectra, the two lines could have slightly different shapes due to the instrument effect and telescope optics. In particular, the [\ion{O}{3}] $\lambda$5007 line is three times stronger than the [\ion{O}{3}] $\lambda$4959 line and thus has a much higher S/N. The combination of the asymmetric line shapes and different S/Ns will likely introduce systematics. Furthermore, line emission from background or foreground objects will likely contaminate our line shapes. In this work we only use narrow and strong emission lines, and we have removed doublet lines with apparently different line shapes. Much higher-resolution observations in the future are needed to investigate these systematics. 

As the DESI survey is ongoing and accumulating more data, we will improve our results as follows.

\begin{itemize}
  \item{We will improve the wavelength calibration by cooperating with the DESI pipeline team. In particular, we will improve the calibration for the wavelength ranges with plenty of sky lines. Currently, we have three steps of the wavelength calibration after the initial step, including the calibration with arc lamp lines and sky lines within the official pipeline and our own wavelength refinement using sky lines. Our hope is to combine the three steps to one and improve the wavelength calibration.}
  
  \item{We will focus on relatively high-redshift ($z\ge0.2$) galaxies. Despite the fact that we have a large number of galaxies with high S/N ratios at low redshift, we do not expect to achieve a sufficiently high accurate wavelength calibration for our purpose, due to the lack of strong sky lines. Strong, isolated (under the DESI resolution) sky lines in the ``R'' arm are also sparse, so we will mainly focus on the redshift range of $z\ge0.4$, particularly some wavelength ranges (redshift slices) with plenty of strong sky lines.}
  
  \item{We will use fainter galaxies. In this pilot study, our galaxies are required to have $Q>5$. We have checked a small sample of fainter galaxies with $3<Q<5$ and $\rm FWHM<250\,km\,s^{-1}$, and found that many of them are quite good. Their spectral quality is generally poorer, meaning larger uncertainties on the $\Delta\alpha/\alpha$ measurements. One problem is that their spectral shapes could be severely affected by strong sky lines, so we need to properly deal with these cases. Note that the number of faint galaxies is much higher than that of brighter galaxies, particularly at $z>0.6$. This is the redshift range for the DESI ELG targets. }
  
  \item{We will use a much larger sample to achieve a more stringent constraint. With the combination of new DESI data and fainter galaxies, we will significantly increase our sample size at $z\ge0.4$. The current uncertainties of $\Delta\alpha/\alpha$ in the $\Delta z =0.1$ bins at $z\ge0.4$ are about $(0.5-1.0)\times 10^{-5}$. We hope to improve them to $(0.2-0.3)\times 10^{-5}$. When all data at $z\ge0.4$ are combined, we expect to reach a better constraint statistically, close to or even comparable to} the strongest constraint from previous quasar observations.

  \item{We will consider more emission lines in the next work, particularly the [\ion{Ne}{3}] doublet emission lines. The [\ion{Ne}{3}] doublet has a wider wavelength separation (nearly 100 \AA), so wavelength calibration is also critical for [\ion{Ne}{3}] to measure $\Delta\alpha/\alpha$. The [\ion{Ne}{3}] emission is generally much weaker than the [\ion{O}{3}] emission in galaxies, and thus we do not expect to obtain a stronger constraint on $\Delta\alpha/\alpha$. Compared to [\ion{O}{3}], however, [\ion{Ne}{3}] allows us to probe higher redshifts (up to $z\approx1.45$). }
  
\end{itemize}

\section{Summary} \label{sec:summary}

We have used a large sample of roughly 110,000 [\ion{O}{3}] emission-line galaxies from DESI to constrain the variation of $\alpha$ in space and time. The galaxy sample was drawn from an internal data release. A quality parameter $Q$ was defined to select strong and narrow [\ion{O}{3}] emitters, since our primary goal is to precisely measure the wavelengths of the [\ion{O}{3}] doublet lines and the wavelength measurement is dependent of both line strength and line width. In this pilot work, we required $Q>5$, and this may be relaxed in the next work. Each emission line was fitted by a single Gaussian or double Gaussian profile and its wavelength was obtained from the best fit. We used the traditional AD method to measure the relative $\alpha$ variation $\Delta\alpha/\alpha$. A great advantage of [\ion{O}{3}] is its wide wavelength separation between the two doublet lines, which makes it very sensitive to the measurement of $\Delta\alpha/\alpha$. On the other hand, this also makes it sensitive to the (relative) wavelength calibration.

Our galaxy sample spans a redshift range of $0<z<0.95$, which covers roughly half of all cosmic time. To estimate the variation in $\alpha$, we grouped the sample into 10 redshift bins with a bin size of $\Delta z = 0.1$. We then calculated $\Delta\alpha/\alpha$ for individual bins using two methods, a weighted average and a bootstrap estimate. The two methods provided well consistent results. The statistical uncertainties of the 10 measured $\Delta\alpha/\alpha$ values are roughly between $2\times10^{-6}$ and $2\times 10^{-5}$. The measured $\Delta\alpha/\alpha$ values show an apparent variation with redshift and the variation amplitude is about $(2\sim3)\times10^{-5}$. This variation should be caused by systematics associated with the wavelength calibration. We found that a small wavelength distortion of $\rm 0.002-0.003\,\AA$ can cause such a variation, and this is beyond the accuracy that the current DESI data can achieve. We have tried to refine the wavelength calibration using sky lines. This can only be done for a small fraction of the galaxies that have strong and isolated sky lines around their [\ion{O}{3}] lines, so it does not change the above results.

The large galaxy sample with a large sky coverage has allowed us to probe the possible variation of $\alpha$ in space. We did it in narrow redshift ranges (Figure \ref{fig:spaceVary}). If the wavelength calibration has a wavelength dependence as we suspected above, this would partially suppress the wavelength dependence. We did not find obvious, large-scale structures in the spatial distribution of $\Delta\alpha/\alpha$ in grids of $\sim15\degr \times 15\degr$. The distribution is quite random at a level of $<10^{-4}$ in individual redshift ranges. 

DESI is still accumulating data, so we expect to have a much larger dataset soon. This work focused on bright galaxies (in terms of the [\ion{O}{3}] emission), but we will consider fainter galaxies in the future. We only used [\ion{O}{3}] in this work, and we will use more emission lines (e.g., [\ion{Ne}{3}]) in the next work. In addition, we plan to improve the wavelength calibration using sky lines. With these improvements, we expect to achieve a better constraint on $\Delta\alpha/\alpha$ that is statistically comparable to the best constraints from previous quasar observations.

\begin{acknowledgments}

We acknowledge support from the National Science Foundation of China (12225301) and the National Key R\&D Program of China (2022YFF0503401).

This research used data obtained with the Dark Energy Spectroscopic Instrument (DESI). DESI construction and operations is managed by the Lawrence Berkeley National Laboratory. This material is based upon work supported by the U.S. Department of Energy, Office of Science, Office of High-Energy Physics, under Contract No. DE–AC02–05CH11231, and by the National Energy Research Scientific Computing Center, a DOE Office of Science User Facility under the same contract. Additional support for DESI was provided by the U.S. National Science Foundation (NSF), Division of Astronomical Sciences under Contract No. AST-0950945 to the NSF’s National Optical-Infrared Astronomy Research Laboratory; the Science and Technology Facilities Council of the United Kingdom; the Gordon and Betty Moore Foundation; the Heising-Simons Foundation; the French Alternative Energies and Atomic Energy Commission (CEA); the National Council of Science and Technology of Mexico (CONACYT); the Ministry of Science and Innovation of Spain (MICINN), and by the DESI Member Institutions: www.desi.lbl.gov/collaborating-institutions. 
The DESI collaboration is honored to be permitted to conduct scientific research on Iolkam Du’ag (Kitt Peak), a mountain with particular significance to the Tohono O’odham Nation. Any opinions, findings, and conclusions or recommendations expressed in this material are those of the author(s) and do not necessarily reflect the views of the U.S. National Science Foundation, the U.S. Department of Energy, or any of the listed funding agencies.

\end{acknowledgments}

\vspace{5mm}
\facilities{KPNO:4m}


\bibliography{ms_alpha}{}

\begin{thebibliography}{}
\expandafter\ifx\csname natexlab\endcsname\relax\def\natexlab#1{#1}\fi
\providecommand{\url}[1]{\href{#1}{#1}}
\providecommand{\dodoi}[1]{doi:~\href{http://doi.org/#1}{\nolinkurl{#1}}}
\providecommand{\doeprint}[1]{\href{http://ascl.net/#1}{\nolinkurl{http://ascl.net/#1}}}
\providecommand{\doarXiv}[1]{\href{https://arxiv.org/abs/#1}{\nolinkurl{https://arxiv.org/abs/#1}}}

\bibitem[{{Abrams} {et~al.}(1994){Abrams}, {Davis}, {Rao}, {Engleman}, \&
  {Brault}}]{Abrams1994}
{Abrams}, M.~C., {Davis}, S.~P., {Rao}, M.~L.~P., {Engleman}, Rolf, J., \&
  {Brault}, J.~W. 1994, \apjs, 93, 351, \dodoi{10.1086/192058}

\bibitem[{{Albareti} {et~al.}(2015){Albareti}, {Comparat}, {Guti{\'e}rrez},
  {Prada}, {P{\^a}ris}, {Schlegel}, {L{\'o}pez-Corredoira}, {Schneider},
  {Manchado}, {Garc{\'\i}a-Hern{\'a}ndez}, {Petitjean}, \& {Ge}}]{Albareti2015}
{Albareti}, F.~D., {Comparat}, J., {Guti{\'e}rrez}, C.~M., {et~al.} 2015,
  \mnras, 452, 4153, \dodoi{10.1093/mnras/stv1406}

\bibitem[{{Alexander} {et~al.}(2023){Alexander}, {Davis}, {Chaussidon},
  {Fawcett}, {X. Gonzalez-Morales}, {Lan}, {Y{\`e}che}, {Ahlen}, {Aguilar},
  {Armengaud}, {Bailey}, {Brooks}, {Cai}, {Canning}, {Carr}, {Chabanier},
  {Cousinou}, {Dawson}, {de la Macorra}, {Dey}, {Dey}, {Dhungana}, {Edge},
  {Eftekharzadeh}, {Fanning}, {Farr}, {Font-Ribera}, {Garcia-Bellido},
  {Garrison}, {Gazta{\~n}aga}, {A Gontcho}, {Gordon}, {Medellin Gonzalez},
  {Guy}, {Herrera-Alcantar}, {Jiang}, {Juneau}, {Kara{\c{c}}ayl{\i}}, {Kehoe},
  {Kisner}, {Kov{\'a}cs}, {Landriau}, {Levi}, {Magneville}, {Martini},
  {Meisner}, {Mezcua}, {Miquel}, {Camacho}, {Moustakas},
  {Mu{\~n}oz-Guti{\'e}rrez}, {Myers}, {Nadathur}, {Napolitano}, {Nie},
  {Palanque-Delabrouille}, {Pan}, {Percival}, {P{\'e}rez-R{\`a}fols},
  {Poppett}, {Prada}, {Ram{\'\i}rez-P{\'e}rez}, {Ravoux}, {Rosario},
  {Schubnell}, {Tarl{\'e}}, {Walther}, {Weiner}, {Youles}, {Zhou}, {Zou}, \&
  {Zou}}]{Alexander2023}
{Alexander}, D.~M., {Davis}, T.~M., {Chaussidon}, E., {et~al.} 2023, \aj, 165,
  124, \dodoi{10.3847/1538-3881/acacfc}

\bibitem[{{Allen}(1973)}]{Allen1973}
{Allen}, C.~W. 1973, {Astrophysical quantities}

\bibitem[{{Alves} {et~al.}(2018){Alves}, {Leite}, {Martins}, {Silva}, {Berge},
  \& {Silva}}]{Alves2018}
{Alves}, C.~S., {Leite}, A.~C.~O., {Martins}, C.~J.~A.~P., {et~al.} 2018, \prd,
  97, 023522, \dodoi{10.1103/PhysRevD.97.023522}

\bibitem[{{Bahcall} \& {Schmidt}(1967)}]{Bahcall1967}
{Bahcall}, J.~N., \& {Schmidt}, M. 1967, \prl, 19, 1294,
  \dodoi{10.1103/PhysRevLett.19.1294}

\bibitem[{{Bahcall} {et~al.}(2004){Bahcall}, {Steinhardt}, \&
  {Schlegel}}]{Bahcall2004}
{Bahcall}, J.~N., {Steinhardt}, C.~L., \& {Schlegel}, D. 2004, \apj, 600, 520,
  \dodoi{10.1086/379971}

\bibitem[{{Barrow} {et~al.}(2002){Barrow}, {Sandvik}, \&
  {Magueijo}}]{Barrow2002}
{Barrow}, J.~D., {Sandvik}, H.~B., \& {Magueijo}, J. 2002, \prd, 65, 063504,
  \dodoi{10.1103/PhysRevD.65.063504}

\bibitem[{{Chand} {et~al.}(2005){Chand}, {Petitjean}, {Srianand}, \&
  {Aracil}}]{Chand2005}
{Chand}, H., {Petitjean}, P., {Srianand}, R., \& {Aracil}, B. 2005, \aap, 430,
  47, \dodoi{10.1051/0004-6361:20041186}

\bibitem[{{Chaussidon} {et~al.}(2023){Chaussidon}, {Y{\`e}che},
  {Palanque-Delabrouille}, {Alexander}, {Yang}, {Ahlen}, {Bailey}, {Brooks},
  {Cai}, {Chabanier}, {Davis}, {Dawson}, {de laMacorra}, {Dey}, {Dey},
  {Eftekharzadeh}, {Eisenstein}, {Fanning}, {Font-Ribera}, {Gazta{\~n}aga}, {A
  Gontcho}, {Gonzalez-Morales}, {Guy}, {Herrera-Alcantar}, {Honscheid},
  {Ishak}, {Jiang}, {Juneau}, {Kehoe}, {Kisner}, {Kov{\'a}cs}, {Kremin}, {Lan},
  {Landriau}, {Le Guillou}, {Levi}, {Magneville}, {Martini}, {Meisner},
  {Moustakas}, {Mu{\~n}oz-Guti{\'e}rrez}, {Myers}, {Newman}, {Nie}, {Percival},
  {Poppett}, {Prada}, {Raichoor}, {Ravoux}, {Ross}, {Schlafly}, {Schlegel},
  {Tan}, {Tarl{\'e}}, {Zhou}, {Zhou}, \& {Zou}}]{Chaussidon2023_DESI_QSO}
{Chaussidon}, E., {Y{\`e}che}, C., {Palanque-Delabrouille}, N., {et~al.} 2023,
  \apj, 944, 107, \dodoi{10.3847/1538-4357/acb3c2}

\bibitem[{{Ciddor}(1996)}]{Ciddor1996}
{Ciddor}, P.~E. 1996, \ao, 35, 1566, \dodoi{10.1364/AO.35.001566}

\bibitem[{{Cooper} {et~al.}(2023){Cooper}, {Koposov}, {Allende Prieto},
  {Manser}, {Kizhuprakkat}, {Myers}, {Dey}, {G{\"a}nsicke}, {Li}, {Rockosi},
  {Valluri}, {Najita}, {Deason}, {Raichoor}, {Wang}, {Ting}, {Kim}, {Carrillo},
  {Wang}, {Beraldo e Silva}, {Han}, {Ding}, {S{\'a}nchez-Conde}, {Aguilar},
  {Ahlen}, {Bailey}, {Belokurov}, {Brooks}, {Cunha}, {Dawson}, {de la Macorra},
  {Doel}, {Eisenstein}, {Fagrelius}, {Fanning}, {Font-Ribera}, {Forero-Romero},
  {Gazta{\~n}aga}, {Gontcho a Gontcho}, {Guy}, {Honscheid}, {Kehoe}, {Kisner},
  {Kremin}, {Landriau}, {Levi}, {Martini}, {Meisner}, {Miquel}, {Moustakas},
  {Nie}, {Palanque-Delabrouille}, {Percival}, {Poppett}, {Prada}, {Rehemtulla},
  {Schlafly}, {Schlegel}, {Schubnell}, {Sharples}, {Tarl{\'e}}, {Wechsler},
  {Weinberg}, {Zhou}, \& {Zou}}]{Cooper2023_DESI_MWS}
{Cooper}, A.~P., {Koposov}, S.~E., {Allende Prieto}, C., {et~al.} 2023, \apj,
  947, 37, \dodoi{10.3847/1538-4357/acb3c0}

\bibitem[{{Cowie} \& {Songaila}(1995)}]{Cowie1995}
{Cowie}, L.~L., \& {Songaila}, A. 1995, \apj, 453, 596, \dodoi{10.1086/176422}

\bibitem[{{Damour} \& {Dyson}(1996)}]{Damour1996}
{Damour}, T., \& {Dyson}, F. 1996, Nuclear Physics B, 480, 37,
  \dodoi{10.1016/S0550-3213(96)00467-1}

\bibitem[{{Dawson} {et~al.}(2013){Dawson}, {Schlegel}, {Ahn}, {Anderson},
  {Aubourg}, {Bailey}, {Barkhouser}, {Bautista}, {Beifiori}, {Berlind},
  {Bhardwaj}, {Bizyaev}, {Blake}, {Blanton}, {Blomqvist}, {Bolton}, {Borde},
  {Bovy}, {Brandt}, {Brewington}, {Brinkmann}, {Brown}, {Brownstein}, {Bundy},
  {Busca}, {Carithers}, {Carnero}, {Carr}, {Chen}, {Comparat}, {Connolly},
  {Cope}, {Croft}, {Cuesta}, {da Costa}, {Davenport}, {Delubac}, {de Putter},
  {Dhital}, {Ealet}, {Ebelke}, {Eisenstein}, {Escoffier}, {Fan}, {Filiz Ak},
  {Finley}, {Font-Ribera}, {G{\'e}nova-Santos}, {Gunn}, {Guo}, {Haggard},
  {Hall}, {Hamilton}, {Harris}, {Harris}, {Ho}, {Hogg}, {Holder}, {Honscheid},
  {Huehnerhoff}, {Jordan}, {Jordan}, {Kauffmann}, {Kazin}, {Kirkby}, {Klaene},
  {Kneib}, {Le Goff}, {Lee}, {Long}, {Loomis}, {Lundgren}, {Lupton}, {Maia},
  {Makler}, {Malanushenko}, {Malanushenko}, {Mandelbaum}, {Manera}, {Maraston},
  {Margala}, {Masters}, {McBride}, {McDonald}, {McGreer}, {McMahon}, {Mena},
  {Miralda-Escud{\'e}}, {Montero-Dorta}, {Montesano}, {Muna}, {Myers},
  {Naugle}, {Nichol}, {Noterdaeme}, {Nuza}, {Olmstead}, {Oravetz}, {Oravetz},
  {Owen}, {Padmanabhan}, {Palanque-Delabrouille}, {Pan}, {Parejko},
  {P{\^a}ris}, {Percival}, {P{\'e}rez-Fournon}, {P{\'e}rez-R{\`a}fols},
  {Petitjean}, {Pfaffenberger}, {Pforr}, {Pieri}, {Prada}, {Price-Whelan},
  {Raddick}, {Rebolo}, {Rich}, {Richards}, {Rockosi}, {Roe}, {Ross}, {Ross},
  {Rossi}, {Rubi{\~n}o-Martin}, {Samushia}, {S{\'a}nchez}, {Sayres}, {Schmidt},
  {Schneider}, {Sc{\'o}ccola}, {Seo}, {Shelden}, {Sheldon}, {Shen}, {Shu},
  {Slosar}, {Smee}, {Snedden}, {Stauffer}, {Steele}, {Strauss}, {Streblyanska},
  {Suzuki}, {Swanson}, {Tal}, {Tanaka}, {Thomas}, {Tinker}, {Tojeiro},
  {Tremonti}, {Vargas Maga{\~n}a}, {Verde}, {Viel}, {Wake}, {Watson}, {Weaver},
  {Weinberg}, {Weiner}, {West}, {White}, {Wood-Vasey}, {Yeche}, {Zehavi},
  {Zhao}, \& {Zheng}}]{Dawson2013}
{Dawson}, K.~S., {Schlegel}, D.~J., {Ahn}, C.~P., {et~al.} 2013, \aj, 145, 10,
  \dodoi{10.1088/0004-6256/145/1/10}

\bibitem[{{Dawson} {et~al.}(2016){Dawson}, {Kneib}, {Percival}, {Alam},
  {Albareti}, {Anderson}, {Armengaud}, {Aubourg}, {Bailey}, {Bautista},
  {Berlind}, {Bershady}, {Beutler}, {Bizyaev}, {Blanton}, {Blomqvist},
  {Bolton}, {Bovy}, {Brandt}, {Brinkmann}, {Brownstein}, {Burtin}, {Busca},
  {Cai}, {Chuang}, {Clerc}, {Comparat}, {Cope}, {Croft}, {Cruz-Gonzalez}, {da
  Costa}, {Cousinou}, {Darling}, {de la Macorra}, {de la Torre}, {Delubac}, {du
  Mas des Bourboux}, {Dwelly}, {Ealet}, {Eisenstein}, {Eracleous}, {Escoffier},
  {Fan}, {Finoguenov}, {Font-Ribera}, {Frinchaboy}, {Gaulme}, {Georgakakis},
  {Green}, {Guo}, {Guy}, {Ho}, {Holder}, {Huehnerhoff}, {Hutchinson}, {Jing},
  {Jullo}, {Kamble}, {Kinemuchi}, {Kirkby}, {Kitaura}, {Klaene}, {Laher},
  {Lang}, {Laurent}, {Le Goff}, {Li}, {Liang}, {Lima}, {Lin}, {Lin}, {Lin},
  {Long}, {Lundgren}, {MacDonald}, {Geimba Maia}, {Malanushenko},
  {Malanushenko}, {Mariappan}, {McBride}, {McGreer}, {M{\'e}nard}, {Merloni},
  {Meza}, {Montero-Dorta}, {Muna}, {Myers}, {Nandra}, {Naugle}, {Newman},
  {Noterdaeme}, {Nugent}, {Ogando}, {Olmstead}, {Oravetz}, {Oravetz},
  {Padmanabhan}, {Palanque-Delabrouille}, {Pan}, {Parejko}, {P{\^a}ris},
  {Peacock}, {Petitjean}, {Pieri}, {Pisani}, {Prada}, {Prakash}, {Raichoor},
  {Reid}, {Rich}, {Ridl}, {Rodriguez-Torres}, {Carnero Rosell}, {Ross},
  {Rossi}, {Ruan}, {Salvato}, {Sayres}, {Schneider}, {Schlegel}, {Seljak},
  {Seo}, {Sesar}, {Shandera}, {Shu}, {Slosar}, {Sobreira}, {Streblyanska},
  {Suzuki}, {Taylor}, {Tao}, {Tinker}, {Tojeiro}, {Vargas-Maga{\~n}a}, {Wang},
  {Weaver}, {Weinberg}, {White}, {Wood-Vasey}, {Yeche}, {Zhai}, {Zhao}, {Zhao},
  {Zheng}, {Ben Zhu}, \& {Zou}}]{Dawson2016}
{Dawson}, K.~S., {Kneib}, J.-P., {Percival}, W.~J., {et~al.} 2016, \aj, 151,
  44, \dodoi{10.3847/0004-6256/151/2/44}

\bibitem[{{de Martino} {et~al.}(2016){de Martino}, {Martins}, {Ebeling}, \&
  {Kocevski}}]{deMartino2016}
{de Martino}, I., {Martins}, C.~J.~A.~P., {Ebeling}, H., \& {Kocevski}, D.
  2016, \prd, 94, 083008, \dodoi{10.1103/PhysRevD.94.083008}

\bibitem[{{DESI Collaboration} {et~al.}(2016{\natexlab{a}}){DESI
  Collaboration}, {Aghamousa}, {Aguilar}, {Ahlen}, {Alam}, {Allen}, {Allende
  Prieto}, {Annis}, {Bailey}, {Balland}, {Ballester}, {Baltay}, {Beaufore},
  {Bebek}, {Beers}, {Bell}, {Bernal}, {Besuner}, {Beutler}, {Blake}, {Bleuler},
  {Blomqvist}, {Blum}, {Bolton}, {Briceno}, {Brooks}, {Brownstein},
  {Buckley-Geer}, {Burden}, {Burtin}, {Busca}, {Cahn}, {Cai}, {Cardiel-Sas},
  {Carlberg}, {Carton}, {Casas}, {Castander}, {Cervantes-Cota}, {Claybaugh},
  {Close}, {Coker}, {Cole}, {Comparat}, {Cooper}, {Cousinou}, {Crocce}, {Cuby},
  {Cunningham}, {Davis}, {Dawson}, {de la Macorra}, {De Vicente}, {Delubac},
  {Derwent}, {Dey}, {Dhungana}, {Ding}, {Doel}, {Duan}, {Ealet}, {Edelstein},
  {Eftekharzadeh}, {Eisenstein}, {Elliott}, {Escoffier}, {Evatt}, {Fagrelius},
  {Fan}, {Fanning}, {Farahi}, {Farihi}, {Favole}, {Feng}, {Fernandez},
  {Findlay}, {Finkbeiner}, {Fitzpatrick}, {Flaugher}, {Flender}, {Font-Ribera},
  {Forero-Romero}, {Fosalba}, {Frenk}, {Fumagalli}, {Gaensicke}, {Gallo},
  {Garcia-Bellido}, {Gaztanaga}, {Pietro Gentile Fusillo}, {Gerard},
  {Gershkovich}, {Giannantonio}, {Gillet}, {Gonzalez-de-Rivera},
  {Gonzalez-Perez}, {Gott}, {Graur}, {Gutierrez}, {Guy}, {Habib}, {Heetderks},
  {Heetderks}, {Heitmann}, {Hellwing}, {Herrera}, {Ho}, {Holland}, {Honscheid},
  {Huff}, {Hutchinson}, {Huterer}, {Hwang}, {Illa Laguna}, {Ishikawa},
  {Jacobs}, {Jeffrey}, {Jelinsky}, {Jennings}, {Jiang}, {Jimenez}, {Johnson},
  {Joyce}, {Jullo}, {Juneau}, {Kama}, {Karcher}, {Karkar}, {Kehoe}, {Kennamer},
  {Kent}, {Kilbinger}, {Kim}, {Kirkby}, {Kisner}, {Kitanidis}, {Kneib},
  {Koposov}, {Kovacs}, {Koyama}, {Kremin}, {Kron}, {Kronig}, {Kueter-Young},
  {Lacey}, {Lafever}, {Lahav}, {Lambert}, {Lampton}, {Landriau}, {Lang},
  {Lauer}, {Le Goff}, {Le Guillou}, {Le Van Suu}, {Lee}, {Lee}, {Leitner},
  {Lesser}, {Levi}, {L'Huillier}, {Li}, {Liang}, {Lin}, {Linder}, {Loebman},
  {Luki{\'c}}, {Ma}, {MacCrann}, {Magneville}, {Makarem}, {Manera}, {Manser},
  {Marshall}, {Martini}, {Massey}, {Matheson}, {McCauley}, {McDonald},
  {McGreer}, {Meisner}, {Metcalfe}, {Miller}, {Miquel}, {Moustakas}, {Myers},
  {Naik}, {Newman}, {Nichol}, {Nicola}, {Nicolati da Costa}, {Nie}, {Niz},
  {Norberg}, {Nord}, {Norman}, {Nugent}, {O'Brien}, {Oh}, {Olsen}, {Padilla},
  {Padmanabhan}, {Padmanabhan}, {Palanque-Delabrouille}, {Palmese},
  {Pappalardo}, {P{\^a}ris}, {Park}, {Patej}, {Peacock}, {Peiris}, {Peng},
  {Percival}, {Perruchot}, {Pieri}, {Pogge}, {Pollack}, {Poppett}, {Prada},
  {Prakash}, {Probst}, {Rabinowitz}, {Raichoor}, {Ree}, {Refregier}, {Regal},
  {Reid}, {Reil}, {Rezaie}, {Rockosi}, {Roe}, {Ronayette}, {Roodman}, {Ross},
  {Ross}, {Rossi}, {Rozo}, {Ruhlmann-Kleider}, {Rykoff}, {Sabiu}, {Samushia},
  {Sanchez}, {Sanchez}, {Schlegel}, {Schneider}, {Schubnell}, {Secroun},
  {Seljak}, {Seo}, {Serrano}, {Shafieloo}, {Shan}, {Sharples}, {Sholl},
  {Shourt}, {Silber}, {Silva}, {Sirk}, {Slosar}, {Smith}, {Smoot}, {Som},
  {Song}, {Sprayberry}, {Staten}, {Stefanik}, {Tarle}, {Sien Tie}, {Tinker},
  {Tojeiro}, {Valdes}, {Valenzuela}, {Valluri}, {Vargas-Magana}, {Verde},
  {Walker}, {Wang}, {Wang}, {Weaver}, {Weaverdyck}, {Wechsler}, {Weinberg},
  {White}, {Yang}, {Yeche}, {Zhang}, {Zhao}, {Zheng}, {Zhou}, {Zhou}, {Zhu},
  {Zou}, \& {Zu}}]{DESI2016a}
{DESI Collaboration}, {Aghamousa}, A., {Aguilar}, J., {et~al.}
  2016{\natexlab{a}}, arXiv e-prints, arXiv:1611.00036.
\newblock \doarXiv{1611.00036}

\bibitem[{{DESI Collaboration} {et~al.}(2016{\natexlab{b}}){DESI
  Collaboration}, {Aghamousa}, {Aguilar}, {Ahlen}, {Alam}, {Allen}, {Allende
  Prieto}, {Annis}, {Bailey}, {Balland}, {Ballester}, {Baltay}, {Beaufore},
  {Bebek}, {Beers}, {Bell}, {Bernal}, {Besuner}, {Beutler}, {Blake}, {Bleuler},
  {Blomqvist}, {Blum}, {Bolton}, {Briceno}, {Brooks}, {Brownstein},
  {Buckley-Geer}, {Burden}, {Burtin}, {Busca}, {Cahn}, {Cai}, {Cardiel-Sas},
  {Carlberg}, {Carton}, {Casas}, {Castander}, {Cervantes-Cota}, {Claybaugh},
  {Close}, {Coker}, {Cole}, {Comparat}, {Cooper}, {Cousinou}, {Crocce}, {Cuby},
  {Cunningham}, {Davis}, {Dawson}, {de la Macorra}, {De Vicente}, {Delubac},
  {Derwent}, {Dey}, {Dhungana}, {Ding}, {Doel}, {Duan}, {Ealet}, {Edelstein},
  {Eftekharzadeh}, {Eisenstein}, {Elliott}, {Escoffier}, {Evatt}, {Fagrelius},
  {Fan}, {Fanning}, {Farahi}, {Farihi}, {Favole}, {Feng}, {Fernandez},
  {Findlay}, {Finkbeiner}, {Fitzpatrick}, {Flaugher}, {Flender}, {Font-Ribera},
  {Forero-Romero}, {Fosalba}, {Frenk}, {Fumagalli}, {Gaensicke}, {Gallo},
  {Garcia-Bellido}, {Gaztanaga}, {Pietro Gentile Fusillo}, {Gerard},
  {Gershkovich}, {Giannantonio}, {Gillet}, {Gonzalez-de-Rivera},
  {Gonzalez-Perez}, {Gott}, {Graur}, {Gutierrez}, {Guy}, {Habib}, {Heetderks},
  {Heetderks}, {Heitmann}, {Hellwing}, {Herrera}, {Ho}, {Holland}, {Honscheid},
  {Huff}, {Hutchinson}, {Huterer}, {Hwang}, {Illa Laguna}, {Ishikawa},
  {Jacobs}, {Jeffrey}, {Jelinsky}, {Jennings}, {Jiang}, {Jimenez}, {Johnson},
  {Joyce}, {Jullo}, {Juneau}, {Kama}, {Karcher}, {Karkar}, {Kehoe}, {Kennamer},
  {Kent}, {Kilbinger}, {Kim}, {Kirkby}, {Kisner}, {Kitanidis}, {Kneib},
  {Koposov}, {Kovacs}, {Koyama}, {Kremin}, {Kron}, {Kronig}, {Kueter-Young},
  {Lacey}, {Lafever}, {Lahav}, {Lambert}, {Lampton}, {Landriau}, {Lang},
  {Lauer}, {Le Goff}, {Le Guillou}, {Le Van Suu}, {Lee}, {Lee}, {Leitner},
  {Lesser}, {Levi}, {L'Huillier}, {Li}, {Liang}, {Lin}, {Linder}, {Loebman},
  {Luki{\'c}}, {Ma}, {MacCrann}, {Magneville}, {Makarem}, {Manera}, {Manser},
  {Marshall}, {Martini}, {Massey}, {Matheson}, {McCauley}, {McDonald},
  {McGreer}, {Meisner}, {Metcalfe}, {Miller}, {Miquel}, {Moustakas}, {Myers},
  {Naik}, {Newman}, {Nichol}, {Nicola}, {Nicolati da Costa}, {Nie}, {Niz},
  {Norberg}, {Nord}, {Norman}, {Nugent}, {O'Brien}, {Oh}, {Olsen}, {Padilla},
  {Padmanabhan}, {Padmanabhan}, {Palanque-Delabrouille}, {Palmese},
  {Pappalardo}, {P{\^a}ris}, {Park}, {Patej}, {Peacock}, {Peiris}, {Peng},
  {Percival}, {Perruchot}, {Pieri}, {Pogge}, {Pollack}, {Poppett}, {Prada},
  {Prakash}, {Probst}, {Rabinowitz}, {Raichoor}, {Ree}, {Refregier}, {Regal},
  {Reid}, {Reil}, {Rezaie}, {Rockosi}, {Roe}, {Ronayette}, {Roodman}, {Ross},
  {Ross}, {Rossi}, {Rozo}, {Ruhlmann-Kleider}, {Rykoff}, {Sabiu}, {Samushia},
  {Sanchez}, {Sanchez}, {Schlegel}, {Schneider}, {Schubnell}, {Secroun},
  {Seljak}, {Seo}, {Serrano}, {Shafieloo}, {Shan}, {Sharples}, {Sholl},
  {Shourt}, {Silber}, {Silva}, {Sirk}, {Slosar}, {Smith}, {Smoot}, {Som},
  {Song}, {Sprayberry}, {Staten}, {Stefanik}, {Tarle}, {Sien Tie}, {Tinker},
  {Tojeiro}, {Valdes}, {Valenzuela}, {Valluri}, {Vargas-Magana}, {Verde},
  {Walker}, {Wang}, {Wang}, {Weaver}, {Weaverdyck}, {Wechsler}, {Weinberg},
  {White}, {Yang}, {Yeche}, {Zhang}, {Zhao}, {Zheng}, {Zhou}, {Zhou}, {Zhu},
  {Zou}, \& {Zu}}]{DESI2016b}
---. 2016{\natexlab{b}}, arXiv e-prints, arXiv:1611.00037.
\newblock \doarXiv{1611.00037}

\bibitem[{{DESI Collaboration} {et~al.}(2022){DESI Collaboration}, {Abareshi},
  {Aguilar}, {Ahlen}, {Alam}, {Alexander}, {Alfarsy}, {Allen}, {Allende
  Prieto}, {Alves}, \& et~al.}]{DESI2022_overview}
{DESI Collaboration}, {Abareshi}, B., {Aguilar}, J., {et~al.} 2022, \aj, 164,
  207, \dodoi{10.3847/1538-3881/ac882b}

\bibitem[{{DESI Collaboration} {et~al.}(2023{\natexlab{a}}){DESI
  Collaboration}, {Adame}, {Aguilar}, {Ahlen}, {Alam}, {Aldering}, {Alexander},
  {Alfarsy}, {Allende Prieto}, {Alvarez}, \& et~al.}]{DESI2023_EDR}
{DESI Collaboration}, {Adame}, A.~G., {Aguilar}, J., {et~al.}
  2023{\natexlab{a}}, arXiv e-prints, arXiv:2306.06308,
  \dodoi{10.48550/arXiv.2306.06308}

\bibitem[{{DESI Collaboration} {et~al.}(2023{\natexlab{b}}){DESI
  Collaboration}, {Adame}, {Aguilar}, {Ahlen}, {Alam}, {Aldering}, {Alexander},
  {Alfarsy}, {Allende Prieto}, {Alvarez}, \& et~al.}]{DESI2023_Validation}
---. 2023{\natexlab{b}}, arXiv e-prints, arXiv:2306.06307,
  \dodoi{10.48550/arXiv.2306.06307}

\bibitem[{{Dey} {et~al.}(2019){Dey}, {Schlegel}, {Lang}, {Blum}, {Burleigh},
  {Fan}, {Findlay}, {Finkbeiner}, {Herrera}, {Juneau}, {Landriau}, {Levi},
  {McGreer}, {Meisner}, {Myers}, {Moustakas}, {Nugent}, {Patej}, {Schlafly},
  {Walker}, {Valdes}, {Weaver}, {Y{\`e}che}, {Zou}, {Zhou}, {Abareshi},
  {Abbott}, {Abolfathi}, {Aguilera}, {Alam}, {Allen}, {Alvarez}, {Annis},
  {Ansarinejad}, {Aubert}, {Beechert}, {Bell}, {BenZvi}, {Beutler}, {Bielby},
  {Bolton}, {Brice{\~n}o}, {Buckley-Geer}, {Butler}, {Calamida}, {Carlberg},
  {Carter}, {Casas}, {Castander}, {Choi}, {Comparat}, {Cukanovaite}, {Delubac},
  {DeVries}, {Dey}, {Dhungana}, {Dickinson}, {Ding}, {Donaldson}, {Duan},
  {Duckworth}, {Eftekharzadeh}, {Eisenstein}, {Etourneau}, {Fagrelius},
  {Farihi}, {Fitzpatrick}, {Font-Ribera}, {Fulmer}, {G{\"a}nsicke},
  {Gaztanaga}, {George}, {Gerdes}, {Gontcho}, {Gorgoni}, {Green}, {Guy},
  {Harmer}, {Hernandez}, {Honscheid}, {Huang}, {James}, {Jannuzi}, {Jiang},
  {Joyce}, {Karcher}, {Karkar}, {Kehoe}, {Kneib}, {Kueter-Young}, {Lan},
  {Lauer}, {Le Guillou}, {Le Van Suu}, {Lee}, {Lesser}, {Perreault Levasseur},
  {Li}, {Mann}, {Marshall}, {Mart{\'\i}nez-V{\'a}zquez}, {Martini}, {du Mas des
  Bourboux}, {McManus}, {Meier}, {M{\'e}nard}, {Metcalfe},
  {Mu{\~n}oz-Guti{\'e}rrez}, {Najita}, {Napier}, {Narayan}, {Newman}, {Nie},
  {Nord}, {Norman}, {Olsen}, {Paat}, {Palanque-Delabrouille}, {Peng},
  {Poppett}, {Poremba}, {Prakash}, {Rabinowitz}, {Raichoor}, {Rezaie},
  {Robertson}, {Roe}, {Ross}, {Ross}, {Rudnick}, {Safonova}, {Saha},
  {S{\'a}nchez}, {Savary}, {Schweiker}, {Scott}, {Seo}, {Shan}, {Silva},
  {Slepian}, {Soto}, {Sprayberry}, {Staten}, {Stillman}, {Stupak}, {Summers},
  {Sien Tie}, {Tirado}, {Vargas-Maga{\~n}a}, {Vivas}, {Wechsler}, {Williams},
  {Yang}, {Yang}, {Yapici}, {Zaritsky}, {Zenteno}, {Zhang}, {Zhang}, {Zhou}, \&
  {Zhou}}]{Dey2019}
{Dey}, A., {Schlegel}, D.~J., {Lang}, D., {et~al.} 2019, \aj, 157, 168,
  \dodoi{10.3847/1538-3881/ab089d}

\bibitem[{{Dumont} \& {Webb}(2017)}]{Dumont2017}
{Dumont}, V., \& {Webb}, J.~K. 2017, \mnras, 468, 1568,
  \dodoi{10.1093/mnras/stx381}

\bibitem[{{Dzuba} {et~al.}(1999{\natexlab{a}}){Dzuba}, {Flambaum}, \&
  {Webb}}]{Dzuba1999PhRvL_82_888}
{Dzuba}, V.~A., {Flambaum}, V.~V., \& {Webb}, J.~K. 1999{\natexlab{a}}, \prl,
  82, 888, \dodoi{10.1103/PhysRevLett.82.888}

\bibitem[{{Dzuba} {et~al.}(1999{\natexlab{b}}){Dzuba}, {Flambaum}, \&
  {Webb}}]{Dzuba1999PhRvA_59_230}
---. 1999{\natexlab{b}}, \pra, 59, 230, \dodoi{10.1103/PhysRevA.59.230}

\bibitem[{Edl\'{e}n(1953)}]{Edlen1953}
Edl\'{e}n, B. 1953, J. Opt. Soc. Am., 43, 339, \dodoi{10.1364/JOSA.43.000339}

\bibitem[{Edlén(1966)}]{Edlen1966}
Edlén, B. 1966, Metrologia, 2, 71, \dodoi{10.1088/0026-1394/2/2/002}

\bibitem[{{Guy} {et~al.}(2023){Guy}, {Bailey}, {Kremin}, {Alam}, {Alexander},
  {Allende Prieto}, {BenZvi}, {Bolton}, {Brooks}, {Chaussidon}, {Cooper},
  {Dawson}, {de la Macorra}, {Dey}, {Dey}, {Dhungana}, {Eisenstein},
  {Font-Ribera}, {Forero-Romero}, {Gazta{\~n}aga}, {Gontcho A Gontcho},
  {Green}, {Honscheid}, {Ishak}, {Kehoe}, {Kirkby}, {Kisner}, {Koposov}, {Lan},
  {Landriau}, {Le Guillou}, {Levi}, {Magneville}, {Manser}, {Martini},
  {Meisner}, {Miquel}, {Moustakas}, {Myers}, {Newman}, {Nie},
  {Palanque-Delabrouille}, {Percival}, {Poppett}, {Prada}, {Raichoor},
  {Ravoux}, {Ross}, {Schlafly}, {Schlegel}, {Schubnell}, {Sharples},
  {Tarl{\'e}}, {Weaver}, {Y{\'e}che}, {Zhou}, {Zhou}, \& {Zou}}]{Guy2023}
{Guy}, J., {Bailey}, S., {Kremin}, A., {et~al.} 2023, \aj, 165, 144,
  \dodoi{10.3847/1538-3881/acb212}

\bibitem[{{Hahn} {et~al.}(2023){Hahn}, {Wilson}, {Ruiz-Macias}, {Cole},
  {Weinberg}, {Moustakas}, {Kremin}, {Tinker}, {Smith}, {Wechsler}, {Ahlen},
  {Alam}, {Bailey}, {Brooks}, {Cooper}, {Davis}, {Dawson}, {Dey}, {Dey},
  {Eftekharzadeh}, {Eisenstein}, {Fanning}, {Forero-Romero}, {Frenk},
  {Gazta{\~n}aga}, {Gontcho A Gontcho}, {Guy}, {Honscheid}, {Ishak}, {Juneau},
  {Kehoe}, {Kisner}, {Lan}, {Landriau}, {Le Guillou}, {Levi}, {Magneville},
  {Martini}, {Meisner}, {Myers}, {Nie}, {Norberg}, {Palanque-Delabrouille},
  {Percival}, {Poppett}, {Prada}, {Raichoor}, {Ross}, {Safonova}, {Saulder},
  {Schlafly}, {Schlegel}, {Sierra-Porta}, {Tarle}, {Weaver}, {Y{\`e}che},
  {Zarrouk}, {Zhou}, {Zhou}, \& {Zou}}]{Hahn2023_DESI_BGS}
{Hahn}, C., {Wilson}, M.~J., {Ruiz-Macias}, O., {et~al.} 2023, \aj, 165, 253,
  \dodoi{10.3847/1538-3881/accff8}

\bibitem[{{Hanuschik}(2003)}]{Hanuschik2003}
{Hanuschik}, R.~W. 2003, \aap, 407, 1157, \dodoi{10.1051/0004-6361:20030885}

\bibitem[{{Hart} \& {Chluba}(2018)}]{Hart2018}
{Hart}, L., \& {Chluba}, J. 2018, \mnras, 474, 1850,
  \dodoi{10.1093/mnras/stx2783}

\bibitem[{{Holanda} {et~al.}(2016){Holanda}, {Busti}, {Cola{\c{c}}o},
  {Alcaniz}, \& {Landau}}]{Holanda2016}
{Holanda}, R.~F.~L., {Busti}, V.~C., {Cola{\c{c}}o}, L.~R., {Alcaniz}, J.~S.,
  \& {Landau}, S.~J. 2016, \jcap, 2016, 055,
  \dodoi{10.1088/1475-7516/2016/08/055}

\bibitem[{{Jiang} {et~al.}(2007){Jiang}, {Fan}, {Vestergaard}, {Kurk},
  {Walter}, {Kelly}, \& {Strauss}}]{Jiang2007}
{Jiang}, L., {Fan}, X., {Vestergaard}, M., {et~al.} 2007, \aj, 134, 1150,
  \dodoi{10.1086/520811}

\bibitem[{{Juneau} {et~al.}(2024){Juneau}, {Canning}, {Alexander}, {Pucha},
  {Fawcett}, {Myers}, {Moustakas}, {Ruiz-Macias}, {Cole}, {Pan}, {Aguilar},
  {Ahlen}, {Alam}, {Bailey}, {Brooks}, {Chaussidon}, {Circosta}, {Claybaugh},
  {Dawson}, {de la Macorra}, {Dey}, {Doel}, {Fanning}, {Forero-Romero},
  {Gazta{\~n}aga}, {Gontcho}, {Gutierrez}, {Hahn}, {Honscheid}, {Kehoe},
  {Kisner}, {Kremin}, {Lambert}, {Landriau}, {Le Guillou}, {Manera}, {Martini},
  {Meisner}, {Miquel}, {Mu{\~n}oz-Guti{\'e}rrez}, {Nie},
  {Palanque-Delabrouille}, {Percival}, {Poppett}, {Prada}, {Ravoux}, {Rezaie},
  {Rossi}, {Sanchez}, {Schlafly}, {Schlegel}, {Schubnell}, {Seo}, {Silber},
  {Siudek}, {Sprayberry}, {Tarl{\'e}}, {Zhou}, \& {Zou}}]{Juneau2024}
{Juneau}, S., {Canning}, R., {Alexander}, D.~M., {et~al.} 2024, arXiv e-prints,
  arXiv:2404.03621, \dodoi{10.48550/arXiv.2404.03621}

\bibitem[{{Kanekar} {et~al.}(2018){Kanekar}, {Ghosh}, \&
  {Chengalur}}]{Kanekar2018}
{Kanekar}, N., {Ghosh}, T., \& {Chengalur}, J.~N. 2018, \prl, 120, 061302,
  \dodoi{10.1103/PhysRevLett.120.061302}

\bibitem[{{King} {et~al.}(2012){King}, {Webb}, {Murphy}, {Flambaum},
  {Carswell}, {Bainbridge}, {Wilczynska}, \& {Koch}}]{King2012}
{King}, J.~A., {Webb}, J.~K., {Murphy}, M.~T., {et~al.} 2012, \mnras, 422,
  3370, \dodoi{10.1111/j.1365-2966.2012.20852.x}

\bibitem[{{Kotu{\v{s}}} {et~al.}(2017){Kotu{\v{s}}}, {Murphy}, \&
  {Carswell}}]{Kotus2017}
{Kotu{\v{s}}}, S.~M., {Murphy}, M.~T., \& {Carswell}, R.~F. 2017, \mnras, 464,
  3679, \dodoi{10.1093/mnras/stw2543}

\bibitem[{{Lan} {et~al.}(2023){Lan}, {Tojeiro}, {Armengaud}, {Prochaska},
  {Davis}, {Alexander}, {Raichoor}, {Zhou}, {Y{\`e}che}, {Balland}, {BenZvi},
  {Berti}, {Canning}, {Carr}, {Chittenden}, {Cole}, {Cousinou}, {Dawson},
  {Dey}, {Douglass}, {Edge}, {Escoffier}, {Glanville}, {A Gontcho}, {Guy},
  {Hahn}, {Howlett}, {Hwang}, {Jiang}, {Kov{\'a}cs}, {Mezcua}, {Moore},
  {Nadathur}, {Oh}, {Parkinson}, {Rocher}, {Ross}, {Ruhlmann-Kleider}, {Sabiu},
  {Said}, {Saulder}, {Sierra-Porta}, {Weiner}, {Yu}, {Zarrouk}, {Zhang}, {Zou},
  {Ahlen}, {Bailey}, {Brooks}, {Cooper}, {de la Macorra}, {Dey}, {Dhungana},
  {Doel}, {Eftekharzadeh}, {Fanning}, {Font-Ribera}, {Garrison},
  {Gazta{\~n}aga}, {Kehoe}, {Kisner}, {Kremin}, {Landriau}, {Le Guillou},
  {Levi}, {Magneville}, {Meisner}, {Miquel}, {Moustakas}, {Myers}, {Newman},
  {Nie}, {Palanque-Delabrouille}, {Percival}, {Poppett}, {Prada}, {Schubnell},
  {Tarl{\'e}}, {Weaver}, {Zhang}, \& {Zhou}}]{Lan2023}
{Lan}, T.-W., {Tojeiro}, R., {Armengaud}, E., {et~al.} 2023, \apj, 943, 68,
  \dodoi{10.3847/1538-4357/aca5fa}

\bibitem[{{Lee} {et~al.}(2023){Lee}, {Webb}, {Carswell}, {Dzuba}, {Flambaum},
  \& {Milakovi{\'c}}}]{Lee2023}
{Lee}, C.-C., {Webb}, J.~K., {Carswell}, R.~F., {et~al.} 2023, \mnras, 521,
  850, \dodoi{10.1093/mnras/stad600}

\bibitem[{{Lee} {et~al.}(2021{\natexlab{a}}){Lee}, {Webb}, {Carswell}, \&
  {Milakovi{\'c}}}]{Lee2021a}
{Lee}, C.-C., {Webb}, J.~K., {Carswell}, R.~F., \& {Milakovi{\'c}}, D.
  2021{\natexlab{a}}, \mnras, 504, 1787, \dodoi{10.1093/mnras/stab977}

\bibitem[{{Lee} {et~al.}(2021{\natexlab{b}}){Lee}, {Webb}, {Milakovi{\'c}}, \&
  {Carswell}}]{Lee2021b}
{Lee}, C.-C., {Webb}, J.~K., {Milakovi{\'c}}, D., \& {Carswell}, R.~F.
  2021{\natexlab{b}}, \mnras, 507, 27, \dodoi{10.1093/mnras/stab2005}

\bibitem[{{Levi} {et~al.}(2013){Levi}, {Bebek}, {Beers}, {Blum}, {Cahn},
  {Eisenstein}, {Flaugher}, {Honscheid}, {Kron}, {Lahav}, {McDonald}, {Roe},
  {Schlegel}, \& {representing the DESI collaboration}}]{Levi2013}
{Levi}, M., {Bebek}, C., {Beers}, T., {et~al.} 2013, arXiv e-prints,
  arXiv:1308.0847.
\newblock \doarXiv{1308.0847}

\bibitem[{{Levshakov} {et~al.}(2017){Levshakov}, {Ng}, {Henkel}, \&
  {Mookerjea}}]{Levshakov2017}
{Levshakov}, S.~A., {Ng}, K.~W., {Henkel}, C., \& {Mookerjea}, B. 2017, \mnras,
  471, 2143, \dodoi{10.1093/mnras/stx1782}

\bibitem[{{Martins}(2017)}]{Martins2017}
{Martins}, C.~J.~A.~P. 2017, Reports on Progress in Physics, 80, 126902,
  \dodoi{10.1088/1361-6633/aa860e}

\bibitem[{{Milakovi{\'c}} \& {Jethwa}(2024)}]{Milakovic2024}
{Milakovi{\'c}}, D., \& {Jethwa}, P. 2024, \aap, 684, A38,
  \dodoi{10.1051/0004-6361/202348532}

\bibitem[{{Milakovi{\'c}} {et~al.}(2021){Milakovi{\'c}}, {Lee}, {Carswell},
  {Webb}, {Molaro}, \& {Pasquini}}]{Milakovic2021}
{Milakovi{\'c}}, D., {Lee}, C.-C., {Carswell}, R.~F., {et~al.} 2021, \mnras,
  500, 1, \dodoi{10.1093/mnras/staa3217}

\bibitem[{{Milakovi{\'c}} {et~al.}(2020){Milakovi{\'c}}, {Pasquini}, {Webb}, \&
  {Lo Curto}}]{Milakovic2020}
{Milakovi{\'c}}, D., {Pasquini}, L., {Webb}, J.~K., \& {Lo Curto}, G. 2020,
  \mnras, 493, 3997, \dodoi{10.1093/mnras/staa356}

\bibitem[{{Molaro} {et~al.}(2013){Molaro}, {Centuri{\'o}n}, {Whitmore},
  {Evans}, {Murphy}, {Agafonova}, {Bonifacio}, {D'Odorico}, {Levshakov},
  {Lopez}, {Martins}, {Petitjean}, {Rahmani}, {Reimers}, {Srianand}, {Vladilo},
  \& {Wendt}}]{Molaro2013}
{Molaro}, P., {Centuri{\'o}n}, M., {Whitmore}, J.~B., {et~al.} 2013, \aap, 555,
  A68, \dodoi{10.1051/0004-6361/201321351}

\bibitem[{{Moustakas} {et~al.}(2023){Moustakas}, {Scholte}, {Dey}, \&
  {Khederlarian}}]{Moustakas2023}
{Moustakas}, J., {Scholte}, D., {Dey}, B., \& {Khederlarian}, A. 2023,
  {FastSpecFit: Fast spectral synthesis and emission-line fitting of DESI
  spectra}, Astrophysics Source Code Library, record ascl:2308.005.
\newblock \doeprint{2308.005}

\bibitem[{{Murphy} {et~al.}(2022){Murphy}, {Berke}, {Liu}, {Flynn}, {Lehmann},
  {Dzuba}, \& {Flambaum}}]{Murphy2022}
{Murphy}, M.~T., {Berke}, D.~A., {Liu}, F., {et~al.} 2022, Science, 378, 634,
  \dodoi{10.1126/science.abi9232}

\bibitem[{{Murphy} {et~al.}(2003){Murphy}, {Webb}, \& {Flambaum}}]{Murphy2003}
{Murphy}, M.~T., {Webb}, J.~K., \& {Flambaum}, V.~V. 2003, \mnras, 345, 609,
  \dodoi{10.1046/j.1365-8711.2003.06970.x}

\bibitem[{{Murphy} {et~al.}(2001{\natexlab{a}}){Murphy}, {Webb}, {Flambaum},
  {Churchill}, \& {Prochaska}}]{Murphy2001_MN_327_1223}
{Murphy}, M.~T., {Webb}, J.~K., {Flambaum}, V.~V., {Churchill}, C.~W., \&
  {Prochaska}, J.~X. 2001{\natexlab{a}}, \mnras, 327, 1223,
  \dodoi{10.1046/j.1365-8711.2001.04841.x}

\bibitem[{{Murphy} {et~al.}(2001{\natexlab{b}}){Murphy}, {Webb}, {Flambaum},
  {Dzuba}, {Churchill}, {Prochaska}, {Barrow}, \&
  {Wolfe}}]{Murphy2001_MN_327_1208}
{Murphy}, M.~T., {Webb}, J.~K., {Flambaum}, V.~V., {et~al.} 2001{\natexlab{b}},
  \mnras, 327, 1208, \dodoi{10.1046/j.1365-8711.2001.04840.x}

\bibitem[{{Murphy} {et~al.}(2001{\natexlab{c}}){Murphy}, {Webb}, {Flambaum},
  {Prochaska}, \& {Wolfe}}]{Murphy2001_MN327_1237}
{Murphy}, M.~T., {Webb}, J.~K., {Flambaum}, V.~V., {Prochaska}, J.~X., \&
  {Wolfe}, A.~M. 2001{\natexlab{c}}, \mnras, 327, 1237,
  \dodoi{10.1046/j.1365-8711.2001.04842.x}

\bibitem[{{Myers} {et~al.}(2023){Myers}, {Moustakas}, {Bailey}, {Weaver},
  {Cooper}, {Forero-Romero}, {Abolfathi}, {Alexander}, {Brooks}, {Chaussidon},
  {Chuang}, {Dawson}, {Dey}, {Dey}, {Dhungana}, {Doel}, {Fanning},
  {Gazta{\~n}aga}, {Gontcho A Gontcho}, {Gonzalez-Morales}, {Hahn},
  {Herrera-Alcantar}, {Honscheid}, {Ishak}, {Karim}, {Kirkby}, {Kisner},
  {Koposov}, {Kremin}, {Lan}, {Landriau}, {Lang}, {Levi}, {Magneville},
  {Napolitano}, {Martini}, {Meisner}, {Newman}, {Palanque-Delabrouille},
  {Percival}, {Poppett}, {Prada}, {Raichoor}, {Ross}, {Schlafly}, {Schlegel},
  {Schubnell}, {Tan}, {Tarle}, {Wilson}, {Y{\`e}che}, {Zhou}, {Zhou}, \&
  {Zou}}]{Myers2023_DESItargets}
{Myers}, A.~D., {Moustakas}, J., {Bailey}, S., {et~al.} 2023, \aj, 165, 50,
  \dodoi{10.3847/1538-3881/aca5f9}

\bibitem[{{Osterbrock} {et~al.}(1997){Osterbrock}, {Fulbright}, \&
  {Bida}}]{Osterbrock1997}
{Osterbrock}, D.~E., {Fulbright}, J.~P., \& {Bida}, T.~A. 1997, \pasp, 109,
  614, \dodoi{10.1086/133920}

\bibitem[{{Osterbrock} {et~al.}(1996){Osterbrock}, {Fulbright}, {Martel},
  {Keane}, {Trager}, \& {Basri}}]{Osterbrock1996}
{Osterbrock}, D.~E., {Fulbright}, J.~P., {Martel}, A.~R., {et~al.} 1996, \pasp,
  108, 277, \dodoi{10.1086/133722}

\bibitem[{{Petrov} {et~al.}(2006){Petrov}, {Nazarov}, {Onegin}, {Petrov}, \&
  {Sakhnovsky}}]{Petrov2006}
{Petrov}, Y.~V., {Nazarov}, A.~I., {Onegin}, M.~S., {Petrov}, V.~Y., \&
  {Sakhnovsky}, E.~G. 2006, \prc, 74, 064610,
  \dodoi{10.1103/PhysRevC.74.064610}

\bibitem[{{Potekhin} \& {Varshalovich}(1994)}]{Potekhin1994}
{Potekhin}, A.~Y., \& {Varshalovich}, D.~A. 1994, \aaps, 104, 89

\bibitem[{{Raichoor} {et~al.}(2023){Raichoor}, {Moustakas}, {Newman}, {Karim},
  {Ahlen}, {Alam}, {Bailey}, {Brooks}, {Dawson}, {de la Macorra}, {de Mattia},
  {Dey}, {Dey}, {Dhungana}, {Eftekharzadeh}, {Eisenstein}, {Fanning},
  {Font-Ribera}, {Garc{\'\i}a-Bellido}, {Gazta{\~n}aga}, {A Gontcho}, {Guy},
  {Honscheid}, {Ishak}, {Kehoe}, {Kisner}, {Kremin}, {Lan}, {Landriau}, {Le
  Guillou}, {Levi}, {Magneville}, {Manera}, {Martini}, {Meisner}, {Myers},
  {Nie}, {Palanque-Delabrouille}, {Percival}, {Poppett}, {Prada}, {Ross},
  {Ruhlmann-Kleider}, {Sabiu}, {Schlafly}, {Schlegel}, {Tarl{\'e}}, {Weaver},
  {Y{\`e}che}, {Zhou}, {Zhou}, \& {Zou}}]{Raichoor2023_DESI_ELG}
{Raichoor}, A., {Moustakas}, J., {Newman}, J.~A., {et~al.} 2023, \aj, 165, 126,
  \dodoi{10.3847/1538-3881/acb213}

\bibitem[{{Rosenband} {et~al.}(2008){Rosenband}, {Hume}, {Schmidt}, {Chou},
  {Brusch}, {Lorini}, {Oskay}, {Drullinger}, {Fortier}, {Stalnaker}, {Diddams},
  {Swann}, {Newbury}, {Itano}, {Wineland}, \& {Bergquist}}]{Rosenband2008}
{Rosenband}, T., {Hume}, D.~B., {Schmidt}, P.~O., {et~al.} 2008, Science, 319,
  1808, \dodoi{10.1126/science.1154622}

\bibitem[{{Rousselot} {et~al.}(2000){Rousselot}, {Lidman}, {Cuby}, {Moreels},
  \& {Monnet}}]{Rousselot2000}
{Rousselot}, P., {Lidman}, C., {Cuby}, J.~G., {Moreels}, G., \& {Monnet}, G.
  2000, \aap, 354, 1134

\bibitem[{{Savedoff}(1956)}]{Savedoff1956}
{Savedoff}, M.~P. 1956, \nat, 178, 688, \dodoi{10.1038/178688b0}

\bibitem[{{Schlafly} {et~al.}(2023){Schlafly}, {Kirkby}, {Schlegel}, {Myers},
  {Raichoor}, {Dawson}, {Aguilar}, {Allende Prieto}, {Bailey}, {BenZvi},
  {Bermejo-Climent}, {Brooks}, {de la Macorra}, {Dey}, {Doel}, {Fanning},
  {Font-Ribera}, {Forero-Romero}, {Garc{\'\i}a-Bellido}, {Gontcho}, {Guy},
  {Hahn}, {Honscheid}, {Ishak}, {Juneau}, {Kehoe}, {Kisner}, {Kremin},
  {Landriau}, {Lang}, {Lasker}, {Levi}, {Magneville}, {Manser}, {Martini},
  {Meisner}, {Miquel}, {Moustakas}, {Newman}, {Nie}, {Palanque-Delabrouille},
  {Percival}, {Poppett}, {Rockosi}, {Ross}, {Rossi}, {Tarl{\'e}}, {Weaver},
  {Y{\`e}che}, \& {Zhou}}]{Schlafly2023}
{Schlafly}, E.~F., {Kirkby}, D., {Schlegel}, D.~J., {et~al.} 2023, arXiv
  e-prints, arXiv:2306.06309, \dodoi{10.48550/arXiv.2306.06309}

\bibitem[{{Schmidt} \& {Bouchy}(2024)}]{Schmidt2024}
{Schmidt}, T.~M., \& {Bouchy}, F. 2024, \mnras, 530, 1252,
  \dodoi{10.1093/mnras/stae920}

\bibitem[{{Schmidt} {et~al.}(2021){Schmidt}, {Molaro}, {Murphy}, {Lovis},
  {Cupani}, {Cristiani}, {Pepe}, {Rebolo}, {Santos}, {Abreu}, {Adibekyan},
  {Alibert}, {Aliverti}, {Allart}, {Allende Prieto}, {Alves}, {Baldini},
  {Broeg}, {Cabral}, {Calderone}, {Cirami}, {Coelho}, {Coretti}, {D'Odorico},
  {Di Marcantonio}, {Ehrenreich}, {Figueira}, {Genoni}, {G{\'e}nova Santos},
  {Gonz{\'a}lez Hern{\'a}ndez}, {Kerber}, {Landoni}, {Leite}, {Lizon}, {Lo
  Curto}, {Manescau}, {Martins}, {Meg{\'e}vand}, {Mehner}, {Micela},
  {Modigliani}, {Monteiro}, {Monteiro}, {Mueller}, {Nunes}, {Oggioni},
  {Oliveira}, {Pariani}, {Pasquini}, {Redaelli}, {Riva}, {Santos}, {Sosnowska},
  {Sousa}, {Sozzetti}, {Su{\'a}rez Mascare{\~n}o}, {Udry}, {Zapatero Osorio},
  \& {Zerbi}}]{Schmidt2021}
{Schmidt}, T.~M., {Molaro}, P., {Murphy}, M.~T., {et~al.} 2021, \aap, 646,
  A144, \dodoi{10.1051/0004-6361/202039345}

\bibitem[{{Smith} {et~al.}(2019){Smith}, {Grin}, {Robinson}, \&
  {Qi}}]{Smith2019}
{Smith}, T.~L., {Grin}, D., {Robinson}, D., \& {Qi}, D. 2019, \prd, 99, 043531,
  \dodoi{10.1103/PhysRevD.99.043531}

\bibitem[{{Songaila} \& {Cowie}(2014)}]{Songaila2014}
{Songaila}, A., \& {Cowie}, L.~L. 2014, \apj, 793, 103,
  \dodoi{10.1088/0004-637X/793/2/103}

\bibitem[{{Tsuzuki} {et~al.}(2006){Tsuzuki}, {Kawara}, {Yoshii}, {Oyabu},
  {Tanab{\'e}}, \& {Matsuoka}}]{Tsuzuki2006}
{Tsuzuki}, Y., {Kawara}, K., {Yoshii}, Y., {et~al.} 2006, \apj, 650, 57,
  \dodoi{10.1086/506376}

\bibitem[{{Uzan}(2003)}]{Uzan2003}
{Uzan}, J.-P. 2003, Reviews of Modern Physics, 75, 403,
  \dodoi{10.1103/RevModPhys.75.403}

\bibitem[{{Uzan}(2011)}]{Uzan2011}
---. 2011, Living Reviews in Relativity, 14, 2, \dodoi{10.12942/lrr-2011-2}

\bibitem[{{Vanden Berk} {et~al.}(2001){Vanden Berk}, {Richards}, {Bauer},
  {Strauss}, {Schneider}, {Heckman}, {York}, {Hall}, {Fan}, {Knapp},
  {Anderson}, {Annis}, {Bahcall}, {Bernardi}, {Briggs}, {Brinkmann}, {Brunner},
  {Burles}, {Carey}, {Castander}, {Connolly}, {Crocker}, {Csabai}, {Doi},
  {Finkbeiner}, {Friedman}, {Frieman}, {Fukugita}, {Gunn}, {Hennessy},
  {Ivezi{\'c}}, {Kent}, {Kunszt}, {Lamb}, {Leger}, {Long}, {Loveday}, {Lupton},
  {Meiksin}, {Merelli}, {Munn}, {Newberg}, {Newcomb}, {Nichol}, {Owen}, {Pier},
  {Pope}, {Rockosi}, {Schlegel}, {Siegmund}, {Smee}, {Snir}, {Stoughton},
  {Stubbs}, {SubbaRao}, {Szalay}, {Szokoly}, {Tremonti}, {Uomoto}, {Waddell},
  {Yanny}, \& {Zheng}}]{Vanden2001}
{Vanden Berk}, D.~E., {Richards}, G.~T., {Bauer}, A., {et~al.} 2001, \aj, 122,
  549, \dodoi{10.1086/321167}

\bibitem[{{Vestergaard} \& {Wilkes}(2001)}]{Vestergaard2001}
{Vestergaard}, M., \& {Wilkes}, B.~J. 2001, \apjs, 134, 1,
  \dodoi{10.1086/320357}

\bibitem[{{Wang} {et~al.}(2022){Wang}, {Jiang}, {Shen}, {Ho}, {Vestergaard},
  {Ba{\~n}ados}, {Willott}, {Wu}, {Zou}, {Yang}, {Wang}, {Fan}, \&
  {Wu}}]{Wang2022}
{Wang}, S., {Jiang}, L., {Shen}, Y., {et~al.} 2022, \apj, 925, 121,
  \dodoi{10.3847/1538-4357/ac3a69}

\bibitem[{{Webb} {et~al.}(1999){Webb}, {Flambaum}, {Churchill}, {Drinkwater},
  \& {Barrow}}]{Webb1999}
{Webb}, J.~K., {Flambaum}, V.~V., {Churchill}, C.~W., {Drinkwater}, M.~J., \&
  {Barrow}, J.~D. 1999, \prl, 82, 884, \dodoi{10.1103/PhysRevLett.82.884}

\bibitem[{{Webb} {et~al.}(2011){Webb}, {King}, {Murphy}, {Flambaum},
  {Carswell}, \& {Bainbridge}}]{Webb2011}
{Webb}, J.~K., {King}, J.~A., {Murphy}, M.~T., {et~al.} 2011, \prl, 107,
  191101, \dodoi{10.1103/PhysRevLett.107.191101}

\bibitem[{{Webb} {et~al.}(2023){Webb}, {Lee}, {Milakovic}, {Flambum}, {Dzuba},
  \& {Magueijo}}]{Webb2024}
{Webb}, J.~K., {Lee}, C.-C., {Milakovic}, D., {et~al.} 2023, arXiv e-prints,
  arXiv:2401.00888, \dodoi{10.48550/arXiv.2401.00888}

\bibitem[{{Webb} {et~al.}(2001){Webb}, {Murphy}, {Flambaum}, {Dzuba}, {Barrow},
  {Churchill}, {Prochaska}, \& {Wolfe}}]{Webb2001}
{Webb}, J.~K., {Murphy}, M.~T., {Flambaum}, V.~V., {et~al.} 2001, \prl, 87,
  091301, \dodoi{10.1103/PhysRevLett.87.091301}

\bibitem[{{Whitmore} \& {Murphy}(2015)}]{Whitmore2015}
{Whitmore}, J.~B., \& {Murphy}, M.~T. 2015, \mnras, 447, 446,
  \dodoi{10.1093/mnras/stu2420}

\bibitem[{{Wilczynska} {et~al.}(2015){Wilczynska}, {Webb}, {King}, {Murphy},
  {Bainbridge}, \& {Flambaum}}]{Wilczynska2015}
{Wilczynska}, M.~R., {Webb}, J.~K., {King}, J.~A., {et~al.} 2015, \mnras, 454,
  3082, \dodoi{10.1093/mnras/stv2148}

\bibitem[{{Zhou} {et~al.}(2023){Zhou}, {Dey}, {Newman}, {Eisenstein}, {Dawson},
  {Bailey}, {Berti}, {Guy}, {Lan}, {Zou}, {Aguilar}, {Ahlen}, {Alam}, {Brooks},
  {de la Macorra}, {Dey}, {Dhungana}, {Fanning}, {Font-Ribera}, {Gontcho},
  {Honscheid}, {Ishak}, {Kisner}, {Kov{\'a}cs}, {Kremin}, {Landriau}, {Levi},
  {Magneville}, {Manera}, {Martini}, {Meisner}, {Miquel}, {Moustakas}, {Myers},
  {Nie}, {Palanque-Delabrouille}, {Percival}, {Poppett}, {Prada}, {Raichoor},
  {Ross}, {Schlafly}, {Schlegel}, {Schubnell}, {Tarl{\'e}}, {Weaver},
  {Wechsler}, {Y{\'e}che}, \& {Zhou}}]{Zhou2023_DESI_LRG}
{Zhou}, R., {Dey}, B., {Newman}, J.~A., {et~al.} 2023, \aj, 165, 58,
  \dodoi{10.3847/1538-3881/aca5fb}

\bibitem[{{Zou} {et~al.}(2017){Zou}, {Zhou}, {Fan}, {Zhang}, {Zhou}, {Nie},
  {Peng}, {McGreer}, {Jiang}, {Dey}, {Fan}, {He}, {Jiang}, {Lang}, {Lesser},
  {Ma}, {Mao}, {Schlegel}, \& {Wang}}]{Zou2017}
{Zou}, H., {Zhou}, X., {Fan}, X., {et~al.} 2017, \pasp, 129, 064101,
  \dodoi{10.1088/1538-3873/aa65ba}

\end{thebibliography}
\bibliographystyle{aasjournal}

\allauthors


\end{document}